\documentclass[prb,aps,twocolumn,floatfix,showpacs,preprintnumbers,amsmath,amssymb,10pt,letterpaper]{revtex4-1}
\usepackage{graphicx}
\usepackage{dcolumn}
\usepackage{bm}
\usepackage{ltxtable}
\usepackage[table]{xcolor}
\usepackage{booktabs}
\usepackage{verbatim}
\usepackage{algorithm}
\usepackage{algorithmic}
\usepackage{ulem}
\begin{document}

\preprint{working draft}

\title{Thermal Equilibration and Thermally-Induced Spin Currents  in a Thin-Film Ferromagnet on a Substrate} 
\author{ Matthew R. Sears } 
\author{ Wayne M. Saslow } 
\email{wsaslow@tamu.edu}
\affiliation{ Department of Physics, Texas A\&M University, College Station, TX 77843-4242 }
\date{\today}

\begin{abstract}
Recent spin-Seebeck experiments on thin ferromagnetic films apply a temperature difference $\Delta T_{x}$ along the length $x$ and measure a (transverse) voltage difference $\Delta V_{y}$ along the width $y$.  The connection between these effects is complex, involving: (1) thermal equilibration between sample and substrate; (2) spin currents along the height (or thickness) $z$; and (3) the measured voltage difference.  The present work studies in detail the first of these steps, and outlines the other two steps.  
Thermal equilibration processes between the magnons and phonons in the sample, as well as between the sample and the substrate leads to two surface modes, with surface lengths $\lambda$, to provide for thermal equilibration.  Increasing the coupling between the two modes increases the longer mode length and decreases the shorter mode length.  The applied thermal gradient along $x$ leads to a thermal gradient along $z$ that varies as $\sinh{(x/\lambda)}$, which can in turn produce fluxes of the carriers of up- and down- spins along $z$, and gradients of their associated \textit{magnetoelectrochemical potentials} $\bar{\mu}_{\uparrow,\downarrow}$, which vary as $\sinh{(x/\lambda)}$.  By the inverse spin Hall effect, this spin current  along $z$ can produce a transverse (along $y$) voltage difference $\Delta V_y$, which also varies as $\sinh{(x/\lambda)}$.  
\end{abstract}



\maketitle

\section{Introduction}  
\begin{figure}[ht!]
\centerline{\includegraphics[width=8.2cm]{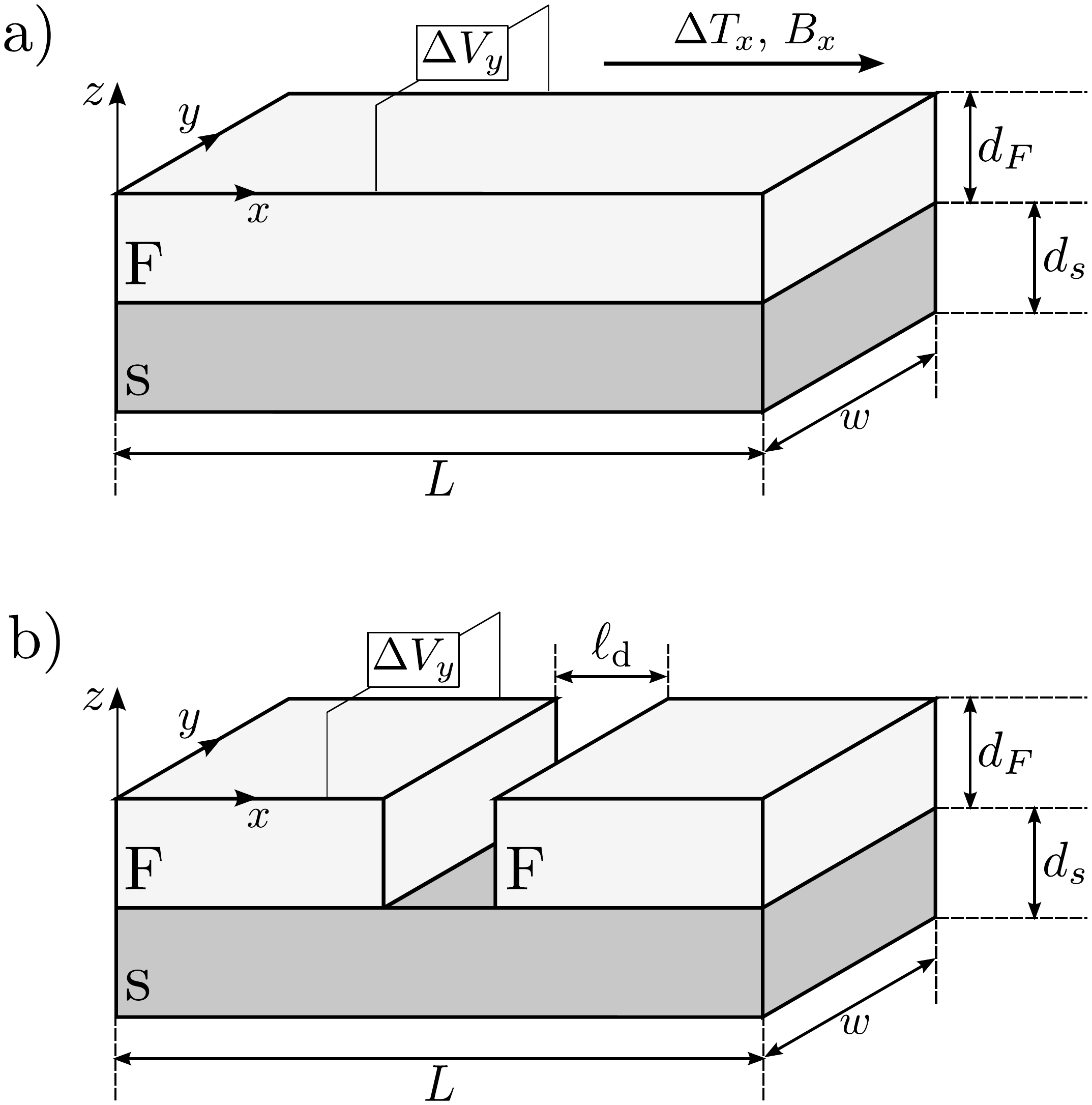}}
\caption{The substrate (s, dark gray) and ferromagnetic sample (F, light gray) of the spin-Seebeck experiment.  Here, (a) shows the typical experimental system, and (b) shows the system with a disconnection (scratch) in the sample (but not the substrate) of length $\ell_d$.  An external magnetic field $B_x$ is applied along $x$, and a temperature difference $\Delta T_x$ along $x$ is maintained by a heater and a heat sink. A voltage difference $\Delta V_{y}$ across the sample in the $y$-direction is measured as a function of $x$ by point electrodes\citep{AwschMyers} or by Pt wires (not shown) deposited on the sample.\citep{AwschMyers,UchidaPy,UchidaInsul}  For a scratch length $\ell_d = 350~\mu$m,  Ref.~\onlinecite{AwschMyers} measures a similar signal $\Delta V_y$ as for the unscratched sample (a).  The figures are not to scale.   The heater and heat sink, which are placed at each edge of the substrate along $x$, are not pictured; see Fig.~\ref{fig:Heat}.}
\label{fig:SSEGeometry} 
\end{figure}

In principle, a thermal gradient $\nabla T$ can produce a spin current.\cite{JohnsonSilsbee}  This magnetic analog of the Seebeck effect, whereby electric currents are generated by $\nabla T$, is known as the spin-Seebeck effect (SSE).  
Evidence for the spin-Seebeck effect has recently been observed in ferromagnet films with thicknesses $d_F \sim 10$~nm and lengtsh $L \sim 10$~mm grown on  insulating substrates.\citep{AwschMyers,UchidaPy,UchidaInsul}  When subjected to a temperature gradient (see Fig.~\ref{fig:SSEGeometry}a), a nonzero voltage difference $\Delta V_y$ across the width of the sample is observed; this signal is attributed to an Inverse Spin Hall Effect (ISHE) due to an inferred spin-Seebeck-induced potential gradient along $z$.  (We employ the magnetoelectrochemical potential $ \bar{\mu}_{\uparrow,\downarrow} $ introduced in Ref.~\onlinecite{JohnsonSilsbee}, and defined in Sec.~\ref{sec:SpinFlux}.)  The magnitude of $\Delta V_y$ is observed to decay in space over a length much greater than a spin-diffusion length.  

The relation between the applied temperature difference and the measured voltage difference is complicated; the connection is represented by
\begin{gather}
\Delta T_x \xrightarrow{\rm Equil.} \partial_z T \xrightarrow{\rm SSE} \partial_z \bar{\mu}_{\uparrow,\downarrow} \xrightarrow{\rm ISHE} \Delta V_y,
\label{LogicalFlow}
\end{gather}
where $\Delta T_x$ is applied and $\Delta V_y$ is measured, and 
``Equil.'' denotes thermal equilibration processes.
The present work shows the details of $\Delta T_x \xrightarrow{\rm Equil.} \partial_z T$, then discusses $\partial_z T \xrightarrow{\rm SSE} \partial_z \bar{\mu}_{\uparrow,\downarrow}$ and $\partial_z \bar{\mu}_{\uparrow,\downarrow} \xrightarrow{\rm ISHE} \Delta V_y$. 

Reference~\onlinecite{AwschMyers} observes the voltage difference $\Delta V_y$ along $y$ to have a $\sinh{(x/\lambda)}$-like form along the sample for some $\lambda=\lambda_{\rm expt}$, thus indicating a surface effect associated with heat input and output.  It has been suggested\citep{Bauer1} that this surface effect is governed by magnon-phonon thermal equilibration\citep{SandWalt} within the sample, which has a characteristic length of $\lambda_{mp}$.  However, Ref.~\onlinecite{Bauer1} argues that  for permalloy (Ni$_{81}$Fe$_{19}$) this equilibration should yield a maximum characteristic length of only $\lambda_{mp}=0.3$~mm, whereas experiment shows the spin-Seebeck effect to have a characteristic length at least an order of magnitude larger.\citep{UchidaPy}  

Further, the effect is unchanged for a large discontinuity\citep{AwschMyers} (of length $\ell_d = 350$~$\mu$m) along $x$ (see Fig.~\ref{fig:SSEGeometry}b); both with and without the discontinuity, a single $\sinh{(x/\lambda)}$ is measured across the entire length $L$ of the system.  Clearly the substrate, which is the only physical connection between the discontinuous regions of the sample, plays an important role.\cite{AwschMyers} 

This work studies heat flow due to the excitations responsible for thermal conduction -- that is, magnons (spin waves) and phonons (lattice vibrations) -- in this system.  We employ irreversible thermodynamics to justify and extend the 1D, two-subsystem approach of Ref.~\onlinecite{SandWalt} to the system of Fig.~\ref{fig:SSEGeometry}, a 2D ($x$ and $z$) system with translational symmetry along $y$, which contains three subsystems: sample phonons (designated by subscript $p$), sample magnons ($m$), and substrate phonons ($s$).   We consider that substrate phonons incident on the interface directly excite only sample phonons, but not magnons (justification for this approximation is discussed below).  
For a non-magnetic sample, the characteristic length of the sample-substrate thermal equilibration satisfies (see below)
\begin{gather}
\lambda_{ps} \sim \sqrt{\frac{\kappa d}{h_K}}.
\label{lamIntro}
\end{gather}
Here, $\kappa$ is a thermal conductivity, $d$ is a length related to the thicknesses of the sample and substrate, and $h_K$ is the thermal boundary conductance.\citep{Pollack,SwartzPohl}

In addition to various geometrical lengths, there are three different lengths associated with Fig.~\ref{fig:SSEGeometry}: the sample magnon-phonon equilibration length $\lambda_{mp}$; the substrate-sample phonon equilibration length $\lambda_{ps}$; and an infinite length $\lambda_{\infty}$ that leads to the usual linear thermal profile.  Recall that Ref.~\onlinecite{AwschMyers} observes a $\sinh{(x/\lambda)}$ profile of the effect.  If $\lambda \ll L$, then $\sinh{(x/\lambda)}$ can decay too close to the boundaries to be experimentally observed.  Conversely, if $\lambda \gg L$, then $\sinh{(x/\lambda)}$ will appear to be linear in $x$, which may explain the linear signal observed by Refs.~\onlinecite{UchidaPy} and \onlinecite{UchidaInsul}.  It is therefore likely that the longer of $\lambda_{ps}$ and $\lambda_{mp}$ is the length observed.  Moreover, because the results are independent of $\ell_{d}$, we expect that the longer length $\lambda_{1} \gg \ell_{d}$ and the shorter length $\lambda_{2} \ll \ell_{d}$.

When both magnon-phonon equilibration (internal to the ferromagnetic sample, and not present for a non-magnetic sample) and sample-substrate equilibration (not present for a sample with no substrate, as in Ref.~\onlinecite{SandWalt}) are present, the coupling between these two modes further separates their characteristic lengths.  That is, the longer length $\lambda_{1}$ and the shorter length $\lambda_{2}$ are respectively greater and less than both $\lambda_{ps}$ and $\lambda_{mp}$.  With $\kappa_m$, $\kappa_p$, and $\kappa_s$ denoting the respective thermal conductivities of magnons in the sample, of phonons in the sample, and of phonons in the substrate, the coupling is given by dimensionless coupling constant that is the product of $R_{mp}$ and $R_{ps}$, which are shown below to be given by
\begin{gather}
R_{mp} \equiv \left(\frac{\kappa_m}{ \kappa_m + \kappa_p}\right) ,\quad 
R_{ps} \equiv   \left( \frac{d_s \kappa_s }{d_{F} \kappa_p + d_s \kappa_s}\right) ,
\label{RmRs}
\end{gather}
The thicknesses $d_s$ and $d_F$ are shown in Fig.~\ref{fig:SSEGeometry}.  
Thus, the 
coupling strength (and the increase of the longer equilibration length) is enhanced via $R_{mp}$ if the magnons account for an appreciable amount of the thermal conductivity of the ferromagnet, and is enhanced via $R_{ps}$ if the substrate is much thicker or has a much larger heat capacity than the ferromagnetic sample.


Section~\ref{sec:Thermo} employs irreversible thermodynamics to find the energy transferred between two systems at different temperatures, specifically considering systems that share a surface (e.g., the sample and substrate) and systems that share a volume (e.g., magnons and phonons in the ferromagnet).  For heat flow only along $x$, Section~\ref{sec:1d} finds the  characteristic lengths of the thermal equilibration modes, as well as the spatial profiles of the phonon and magnon temperatures and heat fluxes.  For heat flow along both $x$ and $z$, Sec.~\ref{sec:2d} finds the shape of the spatial profile of temperatures and heat fluxes, and numerically solves for the characteristic lengths and $z$-dependence of the phonon and magnon heat flux magnitudes.  
Section~\ref{SSElength} compares estimates of the thermal equilibration lengths\cite{Bauer1}  to the observed decay length of $\Delta V_{y}$. 
Section~\ref{sec:SpinFlux} discusses the connection between the thermal gradients found in Sec.~\ref{sec:2d} and the magnetoelectrochemical potentials (which involves the spin-Seebeck effect) and the subsequent connection to $\Delta V_{y}$ (which involves the inverse Spin Hall effect).  Section~\ref{sec:Conclusion} provides a brief summary and conclusion.  Appendix~\ref{App:Details} gives details of the bulk and boundary conditions associated with heat flux along both $x$  and $z$, used in the numerical calculations in Sec.~\ref{sec:2d}.  

It has recently been proposed\citep{AdachiDrag,JaworskiDrag} that electron-phonon drag and magnon-phonon drag processes are important in explaining the results of  Refs.~\onlinecite{UchidaPy,UchidaInsul,AwschMyers}. (The kinetic theory of electron-phonon drag is found, for example, in Refs.~\onlinecite{ZimanElPh}, \onlinecite{GurevichMashkevich89}, and \onlinecite{HannaSondheimer57}.)  This work does not consider such effects.  

\section{Thermodynamics}
\label{sec:Thermo}

Flow described by thermodynamics is properly given by the methods of irreversible thermodynamics.  We present here a derivation of a result central to Ref.~\onlinecite{SandWalt}, which is the basis of Ref.~\onlinecite{Bauer1}, but which is simply written in Ref.~\onlinecite{Keffer}. 

\subsection{General Equilibration of Two Systems}

We consider {\it any} two systems through which heat and entropy (but not matter, quasi-momentum, or momentum) flow.  We later specifically consider energy equilibration between the phonon-magnon subsystems in a ferromagnet (as in Refs.~\onlinecite{Bauer1} and \onlinecite{SandWalt}), as well as energy equilibration between the respective phonon systems of a ferromagnet and a non-magnetic insulator in contact. 

In two such systems, designated $\alpha$ and $\beta$, the energy differentials may be written as
\begin{align}
d E_\alpha = T_\alpha dS_\alpha, \quad d E_\beta = T_\beta dS_\beta,
\end{align}
where $T$ is the temperature and $S$ is the entropy. 
By energy conservation $d E_{\alpha} = - d E_{\beta}$, so
\begin{align}
d S_\alpha = \frac{d E_\alpha}{T_\alpha},\quad d S_\beta = -\frac{d E_\alpha}{T_\beta}.
\end{align}
Since the entropy change must be non-negative,\citep{Callen} we have
\begin{align}
0 \leq \dot{S}_\alpha + \dot{S}_\beta = \left(\frac{1}{T_\alpha} - \frac{1}{T_\beta} \right) \dot{E}_\alpha = \left(\frac{T_\beta - T_\alpha}{T_\alpha T_\beta} \right)\dot{E}_\alpha.
\label{Sdot}
\end{align}
For $\dot{S}_{\alpha}+\dot{S}_{\beta} \geq 0$ to hold we must have
\begin{align}
\dot{E}_\alpha = \zeta  \left(T_\beta-T_\alpha \right),
\label{Edotalpha}
\end{align}
where $\zeta > 0$. 
That is, by irreversible thermodynamics, the energy flux is driven by a difference in intensive thermodynamic quantities.   The proportionality coefficient $\zeta$ has units of a specific heat divided by time, and as noted below depends either on a boundary conductance (for systems that share a common surface) or a relaxation time (for systems that share the same volume). 

Specific heats per unit volume (C) are defined via
\begin{align}
\dot{\varepsilon}_\alpha = C_\alpha \dot{T}_\alpha,\quad \dot{\varepsilon}_\beta = C_\beta \dot{T}_\beta,
\label{dotepsGen}
\end{align}
where $\varepsilon=E/V$ and $V$ is the volume of the system.  
Use of Eqs.~\eqref{Edotalpha} and \eqref{dotepsGen}, and $\dot{E}_{\beta} = - \dot{E}_{\alpha}$, yields
\begin{gather}
\dot{T}_\alpha = \frac{T_\beta - T_\alpha}{\tau_\alpha},\quad \dot{T}_\beta =\frac{T_\alpha - T_\beta}{\tau_\beta},
\label{dotTs}
\end{gather}
where $\tau_\alpha \equiv C_{\alpha} V_{\alpha}/\zeta$ and $\tau_\beta \equiv C_{\beta} V_{\beta}/\zeta$ have units of time.  Then
\begin{align}
\Delta \dot{T}_{\alpha \beta} \equiv \dot{T}_\beta-\dot{T}_\alpha 
=-\frac{T_\beta - T_\alpha}{\tau_{\alpha \beta}} ,
\label{DeltaTdot}
\end{align}
where we define 
\begin{gather}
\tau_{\alpha \beta} \equiv \frac{\tau_{\alpha} \tau_{\beta}}{\tau_{\alpha} + \tau_{\beta}}.
\label{taualphabeta}
\end{gather}
Equation~\eqref{DeltaTdot} justifies Eq.~(1) of Ref.~\onlinecite{SandWalt}.

\subsection{Two Systems Occupying the Same Volume}

Energy conservation in two systems that occupy the same volume $V$ (e.g., the phonon and magnon systems within a ferromagnet) gives $\dot{\varepsilon}_\alpha = - \dot{\varepsilon}_\beta$, so that substitution of Eqs.~\eqref{dotTs} and \eqref{Edotalpha} into Eq.~\eqref{dotepsGen} yields
\begin{gather}
\frac{C_\alpha}{\tau_\alpha} = \frac{C_\beta}{\tau_\beta}=\frac{\zeta}{V}.
\label{CpToCm}
\end{gather}
Then, with $\tau_\beta = (C_\beta/C_\alpha) \tau_\alpha$, equation~\eqref{taualphabeta} gives
\begin{gather}
\frac{C_\alpha}{\tau_\alpha} = \frac{C_\beta}{\tau_\beta} = \left(\frac{C_\alpha C_\beta}{C_\alpha + C_\beta}\right) \tau_{\alpha \beta}^{-1}.
\label{CtaCtb}
\end{gather}
This is the case studied by Ref.~\onlinecite{SandWalt}.

\subsection{Two Systems with a Contact Surface}

For two systems in thermal contact over a surface of area $A$ (e.g., the ferromagnet and substrate's respective phonon systems in Fig.~\ref{fig:SSEGeometry}), 
we write $\zeta= h_K A$,\citep{SwartzPohl,Pollack} so that
\begin{gather}
\dot{E}_\alpha = - \dot{E}_\beta = h_K A \left(T_\beta - T_\alpha \right).
\label{dotEsurf}
\end{gather}
Here $h_K$ is the thermal boundary conductance.  
Substitution of Eqs.~\eqref{dotEsurf} and \eqref{dotTs} into Eq.~\eqref{dotepsGen} gives
\begin{gather}
\tau_{\alpha} = \frac{d_{\alpha} C_{\alpha}}{h_K},\quad \tau_{\beta} = \frac{d_{\beta} C_{\beta}}{h_K},
\end{gather}
where $d$ is the thickness of the material in the direction normal to the contact surface.  Eq.~\eqref{taualphabeta} then gives
\begin{gather}
\tau_{\alpha \beta} = \frac{1}{h_K} \left( \frac{d_\alpha C_\alpha d_\beta C_\beta}{d_\alpha C_\alpha + d_\beta C_\beta} \right).
\end{gather}

\section{Heat Flow in 1d}

\label{sec:1d}

\begin{figure}[t]
\centerline{\includegraphics[width=8.5cm]{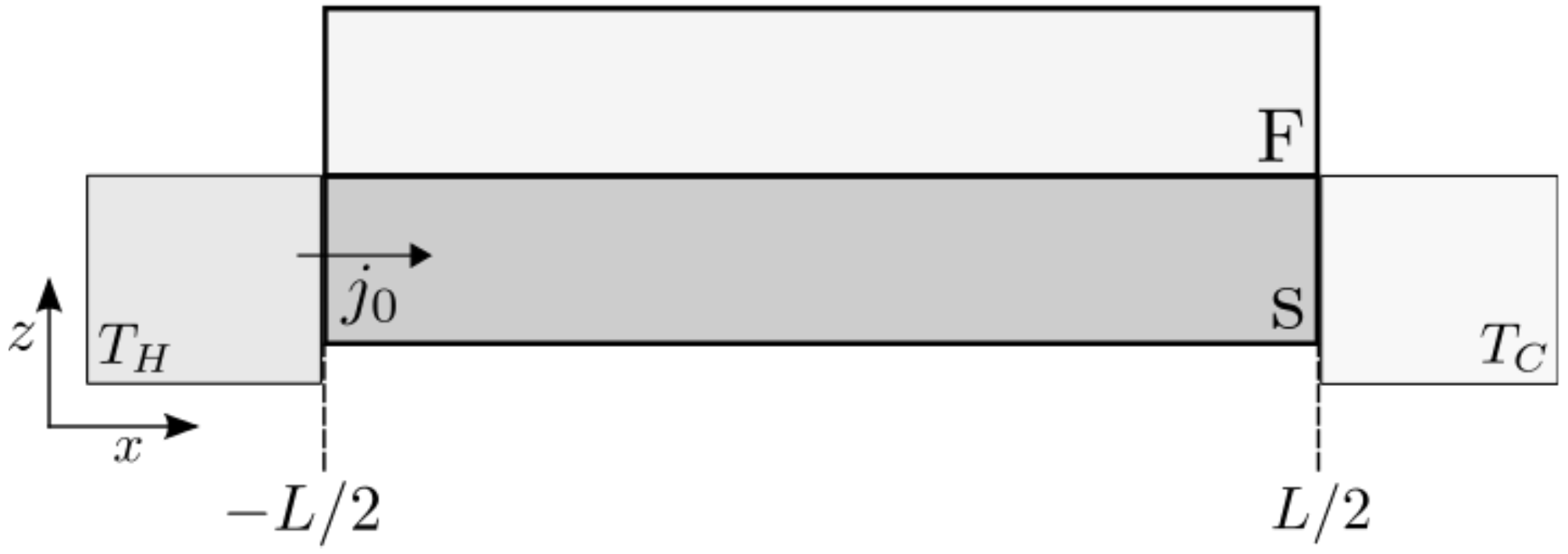}}
\caption{An $xz$-plane cross section of the system under consideration (see Fig.~\ref{fig:SSEGeometry}).  The heater and heat sink, represented by squares at $x<-L/2$ and $x>L/2$, maintain temperatures $T_H$ and $T_C$, where $T_H > T_C$.  For sample isolation, we take them to be in contact only with the substrate (s, dark gray), and not with the ferromagnetic sample (F, light gray); this affects the relative amplitudes of the modes, but not the mode lengths. The total heat flux input by the heater at $x=-L/2$ is $j_{0}$, and a similar heat flux must exit the substrate at $x=L/2$.  In Section~\ref{sec:1d}, we further take all heat fluxes to be uniform in the $yz$-plane; this restriction is lifted in Section~\ref{sec:2d}. }
\label{fig:Heat} 
\end{figure}

We now consider a ferromagnet/substrate system where a thermal gradient is applied by a heater at $x=-L/2$ and a heat sink at $x=L/2$ (see Figure~\ref{fig:Heat}).
For sample isolation, we take them to be in contact only with the substrate.  This affects the relative amplitudes of temperature and thermal flux in each mode, but does not change the mode lengths.

We now take heat to flow only along the length of the materials (the $x$-direction in Figs.~\ref{fig:SSEGeometry} and \ref{fig:Heat}), i.e., heat flow in each system is uniform in the $yz$-plane (Sect~\ref{sec:2d} considers flow along $x$ \textit{and} $z$).  Conservation of energy, with an energy source, is given by
\begin{align}
\dot{\varepsilon} + \partial_x j_x^{\varepsilon} = {\cal S}^{\varepsilon},
\label{EcontGen}
\end{align}
where $j^{\varepsilon}$ is the energy (and heat) flux, and ${\cal S}^{\varepsilon}$ represents the rate of heat transfer per unit volume from one system or subsystem to another.  
We consider steady state solutions, so that $\dot{\varepsilon}=0$. 
Further, we take the magnon system ($m$) in the ferromagnet to only transfer energy to/from the phonon system ($p$) in the ferromagnet.  Similarly we take the substrate ($s$) to only transfer energy to/from the phonon system ($p$) in the ferromagnet, thereby neglecting the magnon-substrate coupling.  

The rate of energy transfer per volume ($V=Ad$) between substrate phonons and sample phonons (an energy source $\cal S$) is found from Eq.~\eqref{dotEsurf} as
\begin{gather}
{\cal S}^{\varepsilon}_{s \rightarrow p} = \frac{h_K}{d_{F}} \left(T_s - T_p\right),\quad 
{\cal S}^{\varepsilon}_{p \rightarrow s} = \frac{h_K}{d_s} \left(T_p - T_s\right).
\end{gather}
Here ${\cal S}^{\varepsilon}_{A\rightarrow B}$ is the volume rate of energy transfer from system $A$ to system $B$.  
This energy transfer is in the form of a source \textit{only} because here we take the heat flux to be only along $x$.  When we include heat flow also along $z$ in Sec.~\ref{sec:2d}, the substrate-sample phonon energy transfer is properly treated as a heat flux along $z$. 

The volume rate of energy transfer between the magnons and phonons in the sample is found by substitution of Eqs.~\eqref{dotTs} and \eqref{CpToCm} into Eq.~\eqref{dotepsGen}, which gives
\begin{gather}
{\cal S}^{\varepsilon}_{m \rightarrow p} = -{\cal S}^{\varepsilon}_{p \rightarrow m} = \frac{C_m}{\tau_m} \left(T_m - T_p\right). 
\label{dotEpdotEm}
\end{gather}
Here we have used Eq.~\eqref{CpToCm} to replace $C_p/\tau_p$ with $C_m/\tau_m$. 
Applied in turn to the substrate, magnons, and phonons, Eq.~\eqref{EcontGen} gives
\begin{gather}
\partial_x j^{\varepsilon_s}_x = \frac{h_K}{d_s} \left(T_p - T_s\right),\label{Econts}\\
\partial_x j^{\varepsilon_m}_x = - \frac{C_m}{\tau_m} \left(T_m - T_p\right),\label{Econtm}\\
\partial_x j^{\varepsilon_p}_x = \frac{h_K}{d_{F}} \left(T_s - T_p\right) + \frac{C_m}{\tau_m}\left(T_m - T_p\right).\label{Econtp}
\end{gather}

As usual, for each subsystem we take the heat flux to be proportional to the gradient of temperature,\citep{Callen,JohnsonSilsbee,SearsSasHeating} so
\begin{gather}
j_i^{\varepsilon} = - \kappa \partial_i T.
\label{jxE}
\end{gather}
Here $\kappa >0$, i.e., heat flows from hot to cold.  
We have neglected cross-terms in Eq.~\eqref{jxE}, where gradients of other intensive thermodynamic quantities also cause a flux; we discuss these cross-terms in further detail in Sec.~\ref{sec:SpinFlux}.  Substitution of Eqs.~\eqref{Econts}, \eqref{Econtm}, and \eqref{Econtp} into the linearized gradient of Eq.~\eqref{jxE} in turn gives
\begin{gather}
-\left(\frac{d_s \kappa_s}{h_K}\right) \partial_x^2 T_s = T_p - T_s,\label{d2Ts}\\
-\left(\frac{\kappa_m \tau_m}{C_m}\right) \partial_x^2 T_m = T_p - T_m,\label{d2Tm}\\
-\kappa_p \partial_x^2 T_p = -\frac{h_K}{d_{F}} \left(T_p - T_s\right) - \frac{C_m}{\tau_m}\left(T_p - T_m\right).\label{d2Tp}
\end{gather} 

\subsection{Characteristic Lengths}
We denote the inhomogeneous parts of $T_s$, $T_p$, and $T_m$ with primes.  They all vary as $e^{\pm q x}$, so the characteristic length is $\lambda = q^{-1}$.  Then, solving  Eqs.~\eqref{d2Ts} and \eqref{d2Tm} for $T_s'$ and $T_m'$ yields
\begin{gather}
T_s' = \frac{T_p'}{\displaystyle 1 -\left(\frac{d_s \kappa_s}{h_K}\right)q^2} ,\quad
T_m' = \frac{T_p'}{\displaystyle 1 - \left(\frac{\kappa_m \tau_m}{C_m}\right)q^2}.\label{TsTmTp}
\end{gather}
Substitution of Eq.~\eqref{TsTmTp} into Eq.~\eqref{d2Tp} gives
\begin{gather}
-\kappa_p q^2 
=\frac{h_K}{d_{F}} \left(\frac{\frac{d_s \kappa_s}{h_K}q^2}{1 -\frac{d_s \kappa_s}{h_K}q^2}\right) + \frac{C_m}{\tau_m} \left(\frac{\frac{\kappa_m \tau_m}{C_m}q^2}{1 - \frac{\kappa_m \tau_m}{C_m}q^2} \right).
\label{qquad}
\end{gather}
This is cubic in $q^2$.  One solution is $q_{\infty}^2=\lambda_{\infty}^{-2}=0$, corresponding to the usual linear temperature profile, for which $T_s'=T_p' = T_m'$. 

We define the inverse lengths $q_{mp}=\lambda_{mp}^{-1}$ and $q_{ps}=\lambda_{ps}^{-1}$, the former associated with magnon-phonon equilibration within the ferromagnet and the latter associated with substrate-sample phonon equilibration.  They satisfy
\begin{gather}
q_{mp}^2 \equiv \frac{C_m}{\tau_m}\left(\frac{\kappa_m+\kappa_p}{\kappa_m \kappa_p}\right),\quad q_{ps}^2 \equiv h_K \left(\frac{d_{F} \kappa_p + d_s \kappa_s}{d_{F} \kappa_p d_s \kappa_s}\right).
\label{qmqsDef}
\end{gather}
They are the inverse lengths of the modes when the magnon-phonon system and the substrate-sample phonon system do not interact.  
Then for $q^2 \neq 0$, equation~\eqref{qquad} can be written as
\begin{gather}
0 = q^4 - q^2 \left(q_{mp}^2+q_{ps}^2\right)+ \left(q_{mp}^2 q_{ps}^2 - q_{mp}^2 q_{ps}^2 R_{mp} R_{ps}\right),
\end{gather}
where the dimensionless ratios $R_{mp}$ and $R_{ps}$ are given by Eq.~\eqref{RmRs}. 
The solutions are
\begin{gather}
q_{(1,2)}^2 = \frac{q_{mp}^2+q_{ps}^2}{2} \pm \sqrt{\left(\frac{q_{mp}^2-q_{ps}^2}{2} \right)^2 +q_{mp}^2 q_{ps}^2 R_{mp} R_{ps} },
\label{qSquare}
\end{gather}
where $q_{1}$ is associated with the minus sign, so that $q_{1}<q_{2}$ and $\lambda_{1}>\lambda_{2}$.

\pagebreak

We now consider two extreme cases.  If there is no substrate (or if $h_K \rightarrow 0$), then
\begin{gather}
|q| \rightarrow  q_{mp} =  \sqrt{\frac{C_m}{\tau_{m}} \left(\frac{\kappa_p + \kappa_m}{\kappa_p \kappa_m} \right)},
\label{qNOs}
\end{gather}
which on use of Eq.~\eqref{CtaCtb} reproduces the result of Ref.~\onlinecite{SandWalt} (which employs $A$ for $q$).  If there is a substrate but no magnons (or $\tau_m \rightarrow \infty$), then
\begin{gather}
|q| \rightarrow  q_{ps} =   \sqrt{h_K \left(\frac{d_s \kappa_s + d_{F} \kappa_p}{d_s \kappa_s d_{F} \kappa_p} \right)} 
,
\label{qNOm}
\end{gather}
as in Eq.~\eqref{lamIntro}.  

\begin{figure}[ht!]
\centerline{\includegraphics[width=8.4cm]{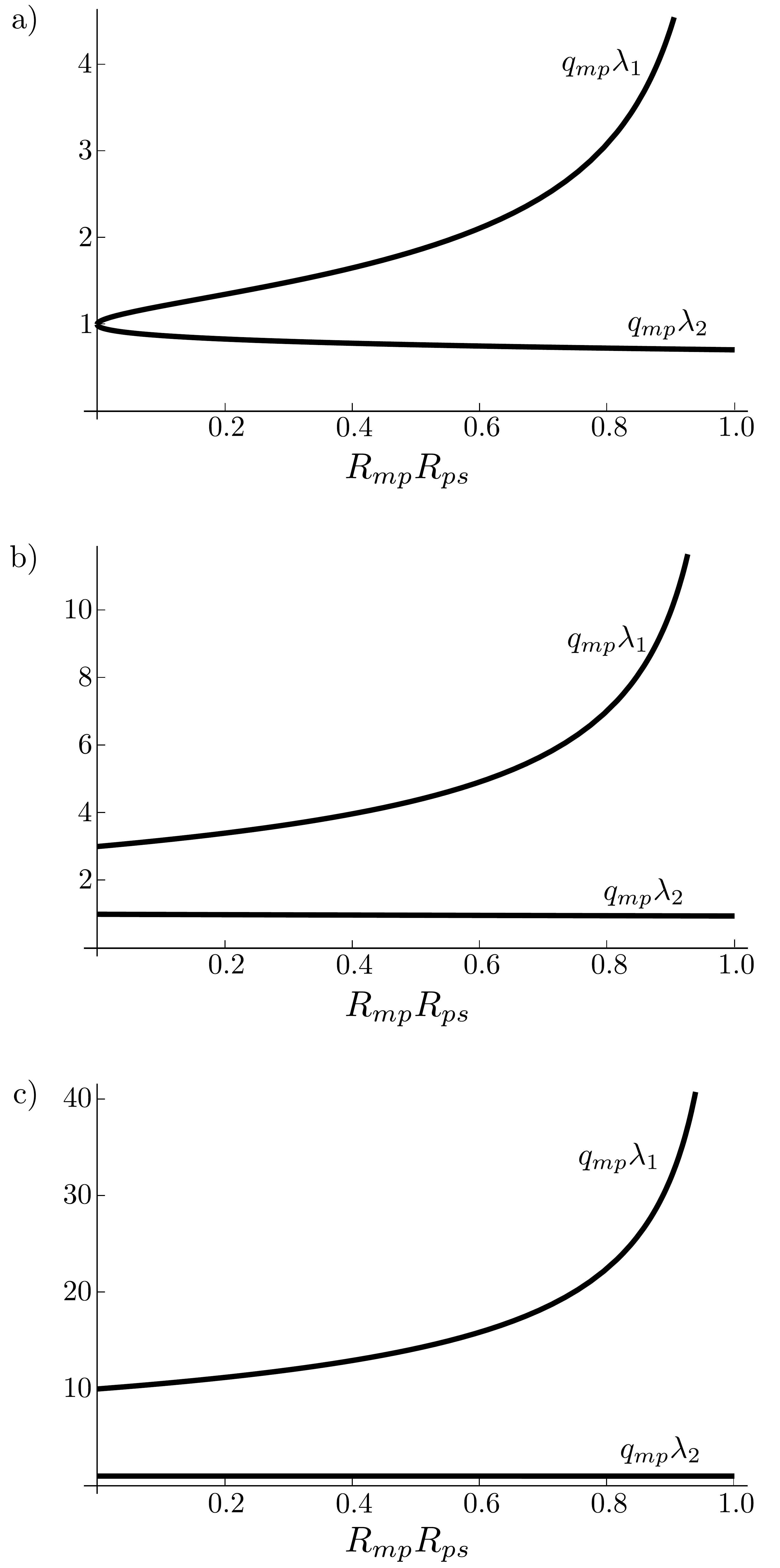}}
\caption{The effect of mode coupling on the characteristic lengths associated with thermal equilibration in the spin-Seebeck system.  The two characteristic lengths $\lambda_{1}$ and $\lambda_{2}$, normalized here by $\lambda_{mp}=q_{mp}^{-1}$, 
are shown as functions of the coupling factor $R_{mp} R_{ps} \leq 1$, for: (a) $q_{mp} = q_{ps}$, which corresponds to equivalent pure mode lengths $\lambda_{mp} = \lambda_{ps}$; (b) $q_{mp} = 3q_{ps}$, which corresponds to $\lambda_{ps} = 3 \lambda_{mp}$; and (c) $q_{mp} = 10 q_{ps}$, which corresponds to $\lambda_{ps} = 10 \lambda_{mp}$.  For $q_{ps} \geq q_{mp}$, the plots are the same when $\lambda_{1}$ and $\lambda_{2}$ are normalized by $\lambda_{ps}$ rather than $\lambda_{mp}$.  By definition, $R_{mp} R_{ps}\leq 1$. 
}
\label{fig:Splitting} 
\end{figure}

The coupling factor ($R_{mp} R_{ps} \leq 1$) between these modes further splits the two solutions; for $R_{mp} R_{ps} \neq 0$, the (shorter) characteristic length $\lambda_{2}=1/q_{2}$ decreases and the (longer) length $\lambda_{1} = 1/q_{1}$ increases.  
For three values of $q_{mp}/q_{ps} \geq 1$, figure~\ref{fig:Splitting} shows the characteristic lengths $\lambda_{1}$ and $\lambda_{2}$, normalized by the pure mode phonon-magnon relaxation length ($\lambda_{mp}=1/q_{mp}$), versus the coupling factor $R_{mp} R_{ps}$.  For $q_{ps} \geq q_{mp}$ the plots are the same when $\lambda_{1}$ and $\lambda_{2}$ are normalized by $q_{ps}$ rather than $q_{mp}$.  

\subsection{Thermal Profile and Fluxes along ${x}$}
We write the phonon temperature in the ferromagnet as
\begin{gather}
T_p = T_0 + \alpha x + \sum_{\gamma=1}^{2} \left[ T^{a}_{\gamma} \sinh{\left( q_{\gamma} x \right)} + T^{b}_{\gamma} \cosh{\left( q_{\gamma} x \right)} \right] , \label{Tp}
\end{gather}
where $T_0$, $T^{a}_{1}$, $T^{a}_{2}$, $T^{b}_{1}$, and $T^{b}_{2}$ are temperatures, and $\alpha$ is a temperature gradient.  The temperatures $T^{a}_{1,2}$ and $T^{b}_{1,2}$ are found by application of the boundary conditions on the heat currents, which are proportional to $\partial_x T_{(p,m,s)}$, with $T^{b}_{1}=0=T^{b}_{2}$ if the heat fluxes have symmetric boundary conditions. 

Recall that $T = T_0 + \alpha x$ for an isolated system under an applied temperature gradient.

Using Eq.~\eqref{qmqsDef}, 
substitution of Eq.~\eqref{Tp} into Eq.~\eqref{TsTmTp} (which applies only to the inhomogeneous parts of $T_{(s,p,m)}$) gives, with no new parameters,
\begin{align}
&T_s = T_0 + \alpha x + \sum_{\gamma=1}^{2} \left[ \frac{q_{ps}^2}{q_{ps}^2 - \left(\frac{d_s \kappa_s + d_{F} \kappa_p}{d_{F} \kappa_p}\right) q_{\gamma}^2} \right]\notag\\
& \qquad \qquad \qquad \times \left[ T^{a}_{\gamma} \sinh{\left( q_{\gamma} x \right)} + T^{b}_{\gamma} \cosh{\left( q_{\gamma} x \right)} \right] ,\label{Ts}\\
&T_m = T_0 + \alpha x + \sum_{\gamma=1}^{2} \left[ \frac{q_{mp}^2}{q_{mp}^2 - \left(\frac{\kappa_m + \kappa_p}{\kappa_p}\right)q_{\gamma}^2}\right] \notag\\
& \qquad  \qquad \qquad \times \left[ T^{a}_{\gamma} \sinh{\left( q_{\gamma} x \right)} + T^{b}_{\gamma} \cosh{\left( q_{\gamma} x \right)} \right].\label{Tm} 
\end{align}
Substituting Eqs.~\eqref{Tp}, \eqref{Ts} and \eqref{Tm} into Eq.~\eqref{jxE} in turn gives the heat current in each subsystem:
\begin{widetext}
\begin{align}
 j_x^{\varepsilon_p} =& -\kappa_p \alpha -\kappa_p \sum_{\gamma=1}^{2} q_{\gamma} \left[ T^{a}_{\gamma} \cosh{\left( q_{\gamma} x \right)} + T^{b}_{\gamma} \sinh{\left( q_{\gamma} x \right)} \right] , \label{jxP}\\
j_x^{\varepsilon_s} =& -\kappa_s \alpha - \kappa_s \sum_{\gamma=1}^{2} q_{\gamma} \left[T^{a}_{\gamma} \cosh{\left( q_{\gamma} x \right)} + T^{b}_{\gamma} \sinh{\left( q_{\gamma} x \right)} \right] \left[\frac{q_{ps}^2  }{q_{ps}^2 - \left(\frac{d_s \kappa_s + d_{F} \kappa_p}{d_{F} \kappa_p}\right)q_{\gamma}^2}\right],\label{jxS}\\
j_x^{\varepsilon_m} =& -\kappa_m \alpha - \kappa_m \sum_{\gamma=1}^{2} q_{\gamma} \left[T^{a}_{\gamma} \cosh{\left( q_{\gamma} x \right)} + T^{b}_{\gamma} \sinh{\left( q_{\gamma} x \right)} \right] \left[\frac{q_{mp}^2 }{q_{mp}^2 - \left(\frac{\kappa_m + \kappa_p}{\kappa_p}\right)q_{\gamma}^2}\right] .\label{jxM}
\end{align} 
\end{widetext}
The total heat flux in the ferromagnet $j_x^{\varepsilon_F} \equiv j_x^{\varepsilon_p}+ j_x^{\varepsilon_m}$ is
\begin{align} 
j_x^{\varepsilon_{F}}=&-\left(\kappa_p + \kappa_m \right)\alpha - \sum_{\gamma=1}^{2} q_{\gamma} \left[\frac{ \left(\kappa_m + \kappa_p \right) \left(q_{mp}^2 - q_{\gamma}^2\right)}{q_{mp}^2 - \left(\frac{\kappa_m + \kappa_p}{\kappa_p}\right) q_{\gamma}^2}\right]\notag\\
& \qquad \qquad \times \left[T^{a}_{\gamma} \cosh{\left( q_{\gamma} x \right)} + T^{b}_{\gamma}  \sinh{\left( q_{\gamma} x \right)} \right].\texttt{}
\end{align}
The boundary conditions on $j_x^{\varepsilon_{(s,p,m)}}$ at $x=-L/2$ and $x=L/2$ give $\alpha$, $T^{a}_{1,2}$ and $T^{b}_{1,2}$.

Because heat flux is continuous, the total heat flux (integrated over all subsystems) due to \textit{each} surface mode must be zero.  This condition is satisfied
by Eqs.~\eqref{jxP}, \eqref{jxS}, and \eqref{jxM} on substitution from Eqs.~\eqref{qmqsDef} and \eqref{qSquare}.

There are five unknowns in Eqs.~\eqref{jxP}, \eqref{jxS}, and \eqref{jxM} ($\alpha$, $T^{a}_{1}$, $T^{a}_{2}$, $T^{b}_{1}$, and $T^{b}_{2}$), and seemingly six boundary conditions (for each of the three fluxes, one at $x=-L/2$ and one at $x=L/2$).  However, 
because the total energy flux is conserved (i.e., no losses at the top of the ferromagnet $d_F$ or at the bottom of the substrate $-d_s$ in Fig.~\ref{fig:SSEGeometry}), there are only five independent conditions.

For comparison to the theory of Ref.~\onlinecite{SandWalt}, we now consider the bulk system if the heaters contact the sample and there is no substrate (so that $q_{2}^2 = q_{mp}^2$ and $q_{1}^2 = 0 = q_{ps}^2$).  Then $j_x^{\varepsilon_{F}} \rightarrow -(\kappa_p+\kappa_m) \alpha$, which reproduces the homogeneous result of Ref.~\onlinecite{SandWalt} (where $Q \equiv j_x^{\varepsilon_{F}}$), and satisfies the condition of zero total heat flux due to the surface mode.  If the heaters directly transfer energy only to and from phonons (so that heat flow in the magnon system vanishes at $x=L/2$ and $x=-L/2$), then $T^a_{2} \rightarrow \kappa_m \alpha/[q_{mp} \kappa_p \cosh{(q_{mp}L/2)}]$ and $T^b_{2}\rightarrow 0$, which reproduces the inhomogeneous solution of Ref.~\onlinecite{SandWalt}.  As noted above, because 
$T_{1,2}^b$ are associated with a term proportional to $\sinh{(q_{1,2} x)}$ in the heat flux, $T_{1}^b=0 = T_{2}^b$ for symmetric boundary conditions on the heat fluxes (i.e., the same heat current is injected into each system at the ``hot" side as is withdrawn from each system at the ``cold" side). 

The above omits any consideration of how heat flows across the sample-substrate interface, which we now address.

\section{Heat flow in 2d}
\label{sec:2d}
We have so far omitted any consideration of how heat flows across the sample-substrate interface, which is now addressed. 
We now consider heat flux along ${z}$, to explicitly permit heat transfer between the substrate and the sample.  We first detail the analytic theory, then present its numerical solution.  

\subsection{Analytic Results}

To completely describe the $z$-dependence of the temperatures and heat fluxes in the system, the $z$-dependence of the heat flux input by the heater at $x=-L/2$ must be considered.  In principle, it may have any functional form, and therefore properly requires a Fourier series in $\sin{(k z)}$ and $\cos{(k z)}$ that includes an infinite number of lengths $k^{-1}$ associated with the $z$-direction.  However, if the thickness (along $z$) of the substrate is much smaller than its length (along $x$), then $k^{-1}$ should be very small compared to $\lambda_{1,2} = q_{1,2}^{-1}$ of Eq.~\eqref{qSquare}.  The contributions from this $z$-dependence should decay along $x$ over a distance on the order of the non-uniformity along $z$, and therefore we do not explicitly include them in the analytic theory.   The cost of neglecting these high $k$ values is that we cannot specify a heat input with a complicated variation along the thickness.



We thus generalize equations~\eqref{Tp}-\eqref{Tm} to take the form
\begin{align}
&T_{(s,p,m)}(x,z) = T_{0_{(s,p,m)}} + \alpha_{(s,p,m)} x \notag\\
&+ \sum_{n=1}^{N} \big[ T^{a}_{(s,p,m){_n}}(z) \sinh{(q_{n} x)} 
 + T^{b}_{(s,p,m){_n}}(z) \cosh{(q_{n} x)} \big].
 \label{Tgen}
\end{align}
Note that we permit there to be $N$ surface modes; for heat flow along only $x$, the one-dimensional heat equations guarantee that $N=2$, but the two-dimensional equations are nonlinear so that any $N$ is allowed.  

The forms of $T^{a}_{(s,p,m){_n}}(z)$ and $T^{b}_{(s,p,m){_n}}(z)$ are determined by the conditions on the heat flux.  
We take symmetric boundary conditions on heat flux along $x$, which give $T^{b}_{(s,p,m){_n}}(z)=0$.
Then, substitution of Eq.~\eqref{Tgen} into Eq.~\eqref{jxE} gives the heat fluxes along $x$ and $z$ to be
\begin{align}
j_x^{\varepsilon_{(s,p,m)}} =& -\kappa_{(s,p,m)} \alpha_{(s,p,m)} \notag\\
&- \kappa_{(s,p,m)} \sum_{n=1}^{N} q_{n} T^{a}_{(s,p,m){_n}}(z) \cosh{(q_{n} x )} ,\label{jxGen}\\
j_z^{\varepsilon_{(s,p,m)}} =& - \kappa_{(s,p,m)} \sum_{n=1}^{N} \partial_z T^{a}_{(s,p,m){_n}}(z) \sinh{(q_{n} x )}.\label{jzGen}
\end{align}
This section finds the functional forms of $T^{a}_{(s,p,m){_n}}(z)$ and shows their amplitudes for example material parameters.  It also discusses the bulk and boundary conditions that permit determination of their amplitudes, with the details of these conditions given by Appendix~\ref{App:Details}. 
 
On properly treating the heat transfer between sample phonons and substrate phonons as $z$-directional currents, and heat transfer between sample magnons and sample phonons as a source/sink as for the 1D case, employing Eqs.~\eqref{jxE} and \eqref{EcontGen} gives
\begin{gather}
\partial_i^2 T_s = 0, \label{d2Tsz}\\
-\kappa_p \partial_i^2 T_p = \frac{C_m}{\tau_m} (T_m - T_p),\label{d2Tpz}\\
-\kappa_m \partial_i^2 T_m = -\frac{C_m}{\tau_m} (T_m - T_p).\label{d2Tmz}
\end{gather}
Equations~\eqref{d2Tsz}-\eqref{d2Tmz} are identical to Eqs.~\eqref{d2Ts}-\eqref{d2Tp}, but with phonon-substrate heat transfer in the form of fluxes rather than sources.  
These equations give
\begin{gather}
 T_{0_m} = T_{0_p} \equiv T_0,\qquad \alpha_{m} = \alpha_{p} \equiv \alpha, 
 \label{T0alpha}
\end{gather}
but they do not explicitly impose any conditions on $T_{0_s}$ or $\alpha_s$.  For steady-state flow, however, we must take 
\begin{gather}
T_{0_s} = T_0, \qquad \alpha_s = \alpha.
\label{T0alphas}
\end{gather}
This relation guarantees that for any two of $\kappa_{(s,p,m)}$ to go continuously to zero, we recover the expected $j_x^{\varepsilon}=-\kappa \alpha$. 
We now find $T^{a}_{(s,p,m){_n}}(z)$ by substituting Eq.~\eqref{Tgen} into Eq.~\eqref{d2Tsz} and the decoupled forms of Eqs.~\eqref{d2Tpz} and \eqref{d2Tmz}. 

Substitution of Eq.~\eqref{Tgen} into Eq.~\eqref{d2Tsz} gives 
\begin{gather}
\partial_z^2 T^{a}_{s{_n}}(z) = - q_{n}^2 T^{a}_{s{_n}}(z),
\label{d2Gs}
\end{gather}
so that $T^{a}_{s{_n}}(z)$ is sinusoidal:
\begin{gather}
T^{a}_{s{_n}}(z) = A_{s{_n}}^{(1)} \cos{(q_{n} z)} + A_{s{_n}}^{(2)} \sin{(q_{n} z)}.
\label{Gs}
\end{gather}  
Here, $A_{s{_n}}^{(1)}$ and $A_{s{_n}}^{(2)}$ are constants determined by conditions on heat flux (see Appendix~\ref{App:Details}).

Decoupled equations for $T_p$ and $T_m$, and thus for  $T^{a}_{p{_n}}(z)$ and $T^{a}_{m{_n}}(z)$, are found by combination of Eqs.~\eqref{d2Tpz} and \eqref{d2Tmz}.  Addition and subtraction gives
\begin{gather}
-\kappa_p \partial_i^2 T_p - \kappa_m \partial_i^2 T_m = 0,\label{d2TpTm1}\\
-\kappa_p \partial_i^2 T_p + \kappa_m \partial_i^2 T_m = 2 \frac{C_m}{\tau_m} (T_m - T_p).\label{d2TpTm2}
\end{gather}
Combination of Eqs.~\eqref{d2TpTm1} and \eqref{d2TpTm2} gives
\begin{gather}
\partial_i^2 \partial_j^2 T_p - q_{mp}^2 \partial_i^2 T_p = 0,
\label{BiharmTp}\\
\partial_i^2 \partial_j^2 T_m - q_{mp}^2 \partial_i^2 T_m = 0.
\label{BiharmTm}
\end{gather}
where we have employed Eq.~\eqref{qmqsDef}. 
Use of Eq.~\eqref{Tgen} in Eqs.~\eqref{BiharmTp} and \eqref{BiharmTm} gives, for each mode $n$,
\begin{gather}
\partial_z^4 T^{a}_{(p,m){_n}}(z) + q_{n}^4 T^{a}_{(p,m){_n}}(z) + 2 q_{n}^2 \partial_z^2 T^{a}_{(p,m){_n}}(z)\notag\\
 - q_{mp}^2 \partial_z^2 T^{a}_{(p,m){_n}}(z) - q_{mp}^2 q_{n}^2 T^{a}_{(p,m){_n}}(z) = 0.
 \label{BiharmG}
\end{gather}
The solution of Eq.~\eqref{BiharmG} is
\begin{align}
T^{a}_{(p,m){_n}}(z) &= A_{(p,m){_n}}^{(1)} e^{\sqrt{q_{mp}^2 - q_{n}^2} z} + A_{(p,m){_n}}^{(2)} e^{-\sqrt{q_{mp}^2 - q_{n}^2} z} \notag\\
&\,+ A_{(p,m){_n}}^{(3)} \cos{(q_{n}z)} + A_{(p,m){_n}}^{(4)} \sin{(q_{n}z)} .
\label{Ggen}
\end{align}
Here, $A_{(p,m){_n}}^{(1,2,3,4)}$ are constants determined by conditions on heat flux (see Appendix~\ref{App:Details}).

Due to the mode splitting discussed in Sec.~\ref{sec:1d}, the 1D inverse lengths straddle $q_{mp}$, that is, $q_{2}^{\rm (1D)} \geq  q_{mp} \geq q_{1}^{\rm (1D)}$.  Therefore, for coupling $R_{mp} R_{ps} \neq 0$, the exponential terms in Eq.~\eqref{Ggen} are, in fact, oscillating terms for each mode that has $q_{n} \gtrsim q_{2}^{\rm (1D)} $.  


\subsection{Bulk and Boundary Conditions}

Although $T_0$, $\alpha$, $A_{s{_n}}^{(1,2)}$, and $A_{(p,m){_n}}^{(1,2,3,4)}$ are 2 + 10$N$ unknowns associated with the temperatures and heat fluxes, they are not free parameters.  
As shown in Appendix~\ref{App:Details}, bulk energy conservation gives 4$N$ conditions; energy conservation at the boundaries $z=-d_s$ and $z=d_F$, where we assume no heat loss to the vacuum, gives 3$N$ conditions; there are $2N$ conditions on heat flux at the interface $z=0$; and there are $2+N$ conditions on temperature and heat flux near the boundaries $x=\pm L/2$.  With these conditions, the present theory has no fitting parameters.  

Specifically, the $3N$ boundary conditions at $z=-d_s$ and $z=d_F$ are given by 
\begin{gather}
j_z^{\varepsilon_m}(x,z=d_F) = 0,\label{jzmATdf}\\
j_z^{\varepsilon_p}(x,z=d_F) = 0,\label{jzpATdf}\\
j_z^{\varepsilon_s}(x,z=-d_s) = 0.\label{jzsATds}
\end{gather}
As discussed in Refs.~\onlinecite{SwartzPohl,Pollack,SearsSasHeating}, heat currents are driven across an interface by the temperature difference across the interface, so that
\begin{gather}
j_z^{\varepsilon_s}(x,z=0) = - h_K \left[ T_p(x,z=0)-T_s(x,z=0) \right],\label{jzsAT0}
\end{gather}  
which gives $N$ conditions.  At the interface we take heat to be transferred only between substrate and sample phonon systems, so that
\begin{gather}
j_z^{\varepsilon_p}(x,z=0) = j_z^{\varepsilon_s}(x,z=0)\label{jzpAT0},
\end{gather}  
or equivalently
\begin{gather}
j_z^{\varepsilon_m}(x,z=0) = 0 \label{jzmAT0},
\end{gather}
giving another $N$ conditions.  
One imposes any two of Eqs.~\eqref{jzsAT0}, \eqref{jzpAT0}, and \eqref{jzmAT0}, with the third being implicitly guaranteed by the energy conservation in the equations of motion.  

Only the remaining conditions, associated with the boundaries $x=-L/2$ and $x=L/2$, can be varied: the average temperature $T_0$, the temperature gradient $\alpha$, and one condition per mode, associated with the relative amounts of heat carried by each subsystem at a given short distance from the heater.  All of these $2+N$ conditions are set by experiment, the first two of which are, respectively, proportional to the sum and difference of the heater and heat sink temperatures.  The other $N$ conditions are related to the relative amounts of heat flux carried by each subsystem near the heater or heat sink.  These conditions are non-obvious, but Appendix~\ref{App:Details} argues that they may be approximated by assuming that near the heater the heat flux carried along $x$ by the substrate phonons dominates that carried by either the sample phonons or sample magnons.  




\subsection{Numerical Solution}
One can not assume that the inverse lengths for 1D heat flow, given by Eq.~\eqref{qSquare} and now called $q_{1}^{\rm (1D)}$ and $q_{2}^{\rm (1D)}$, 
are equivalent to the inverse lengths associated with 2D flow.  Indeed, numerical solution with either of $q_{1}^{\rm (1D)}$ or $q_{2}^{\rm (1D)}$ can be shown to be inconsistent with energy conservation.  Since the 2D heat flow equations are nonlinear, analytic solution is not possible in general.  However, an iterative approach 
can be used to find consistent values for $q$: solve the appropriate boundary conditions for the mode amplitude coefficients (i.e.,  the coefficients $A_{(s,p,m){_n}}^{(k)}$ in Eqs.~\eqref{Gs} and \eqref{Ggen}) using $q_{\rm init} = q_{1}^{\rm (1D)}$ or $q_{\rm init}= q_{2}^{\rm (1D)}$; using these values for the coefficients, find the $q_{\rm new}$ that guarantees energy conservation; begin the loop again using an appropriately chosen $q_{\rm init}'$ in between $q_{\rm init}$ and $q_{\rm new}$.  One must iterate until $q_{\rm new}$ and $q_{\rm init}$ converge.\footnote{Care must be taken in determining a new initial value for the next iteration.  For $q_{\rm init}$ far from a consistent value (that is, a value that satisfies energy conservation), then $q_{\rm init}$ and $q_{\rm new}$ will differ significantly.  Naively choosing $q_{\rm init}' = q_{\rm init} + \frac{1}{2} (q_{\rm new} - q_{\rm init})$ can result in a non-converging series.  Hence, we include the factor $1/C$ in place of $1/2$ to define $q'_{\rm init}$; depending on the initial choice of $q_{\rm init}$, convergence can require $C \sim 10^5$ or greater.}  


For our numerical calculations, we use the material parameters given in Table~\ref{tab:Jaworski}.  Note that Ref.~\onlinecite{Bauer1} estimates $\lambda_{mp}$ to be at least an order of magnitude too small to be the unusually large decay length of the observed voltage difference $\Delta V_{y}$, and the present theory does not explain such a large discrepancy, because as shown in Fig.~\ref{fig:Splitting}, we do not predict mode coupling to amplify the larger length by a full order of magnitude.  This matter is discussed further below.  For the numerical solution, we therefore estimate $\lambda_{mp}= 2$~mm from the observed voltage decay length in Fig.~2 of Ref.~\onlinecite{AwschMyers}.  
We now present the results of this method, calculated using Mathematica~v.~8.0.  

\begin{table}[h!t]
\begin{center}
\caption[Parameters used for numerical calculations of normal mode contributions to the heat fluxes in the spin-Seebeck system.]
{Parameters used in numerical calculations, results of which are shown in Fig.~\ref{fig:jz3D}.  
$^{(a)}$Taken from Fig.~3 of Ref.~\onlinecite{JaworskiDrag}.  
$^{(b)}$To our knowledge, this has not been measured, so we make an order of magnitude estimation.  
$^{(c)}$Value unknown; $\kappa_m/\kappa_p$ is likely to be lower at high temperature.  
$^{(d)}$Estimate from Fig.~2 of Ref.~\onlinecite{AwschMyers} for the decay length of the observed spin-Seebeck voltage signal. 
$^{(e)}$Estimate for Rh:Fe on Al$_2$O$_3$ from Fig.~34 of Ref.~\onlinecite{SwartzPohl}. 
}
\begin{tabularx}{0.45\textwidth}{l r X l l}
\specialrule{1pt}{8pt}{2pt}
\specialrule{1pt}{0pt}{4pt}
Parameter& Value && Units & Ref. \\ 
\specialrule{0.75pt}{4pt}{4pt}
$\kappa_s $ \hspace*{42pt} & 500 && W/m-K  \hspace*{12pt} & \onlinecite{JaworskiDrag}$^{(a)}$\\ [1ex]
$\kappa_p $ & 100 && W/m-K \hspace*{22pt}&  ${ }^{(b)}$ \\ [1ex]
$\kappa_m/\kappa_p$ & 1/10 && & ${ }^{(c)}$ \\ [1ex]
$d_F$ &  $1 \times 10^{-7}$ && m  & \onlinecite{JaworskiDrag} \\  [1ex]
$d_s$ &  $5 \times 10^{-4}$ && m & \onlinecite{JaworskiDrag} \\   [1ex]
$q_{mp}$ & $5 \times 10^2$ && m$^{-1}$ &\onlinecite{AwschMyers}$^{(d)}$\\  [1ex]
$h_K$ & $1 \times 10^{7}$ && W/m$^2$-K & \onlinecite{SwartzPohl}${}^{(e)}$ \\ [1ex]
$L$ & 15.5$\times 10^{-3}$ && m & \onlinecite{JaworskiDrag}\\  [1ex]
\specialrule{1pt}{4pt}{0pt}
\specialrule{1pt}{2pt}{8pt}
\end{tabularx}
\label{tab:Jaworski}
\end{center}
\end{table}

Following Table~\ref{tab:Jaworski}, Eq.~\eqref{qSquare} gives
\begin{align}
&q_{1}^{\rm (1D)} = 476.73~{\rm m}^{-1}, \quad 
 q_{2}^{\rm (1D)} = 1.0000 \times 10^{6}~{\rm m}^{-1}.
\end{align}
Using these as trial values for the numerical solution of 2D heat flow boundary conditions, we find 2D inverse lengths consistent with energy conservation to be
\begin{gather}
q_{1}^{\rm (2D)}= 476.73~{\rm m}^{-1}, \quad q_{2}^{\rm (2D)} = 1.0015 \times 10^{6}~{\rm m}^{-1}.
\end{gather}
Although $q_{1}^{\rm (1D)}$ and $q_{1}^{\rm (2D)}$ match to one part in $10^{8}$ (not shown to this precision above), only $q_{1}^{\rm (2D)}$ satisfies energy conservation. 

\begin{figure}[ht!]
\centerline{\includegraphics[width=0.46\textwidth]{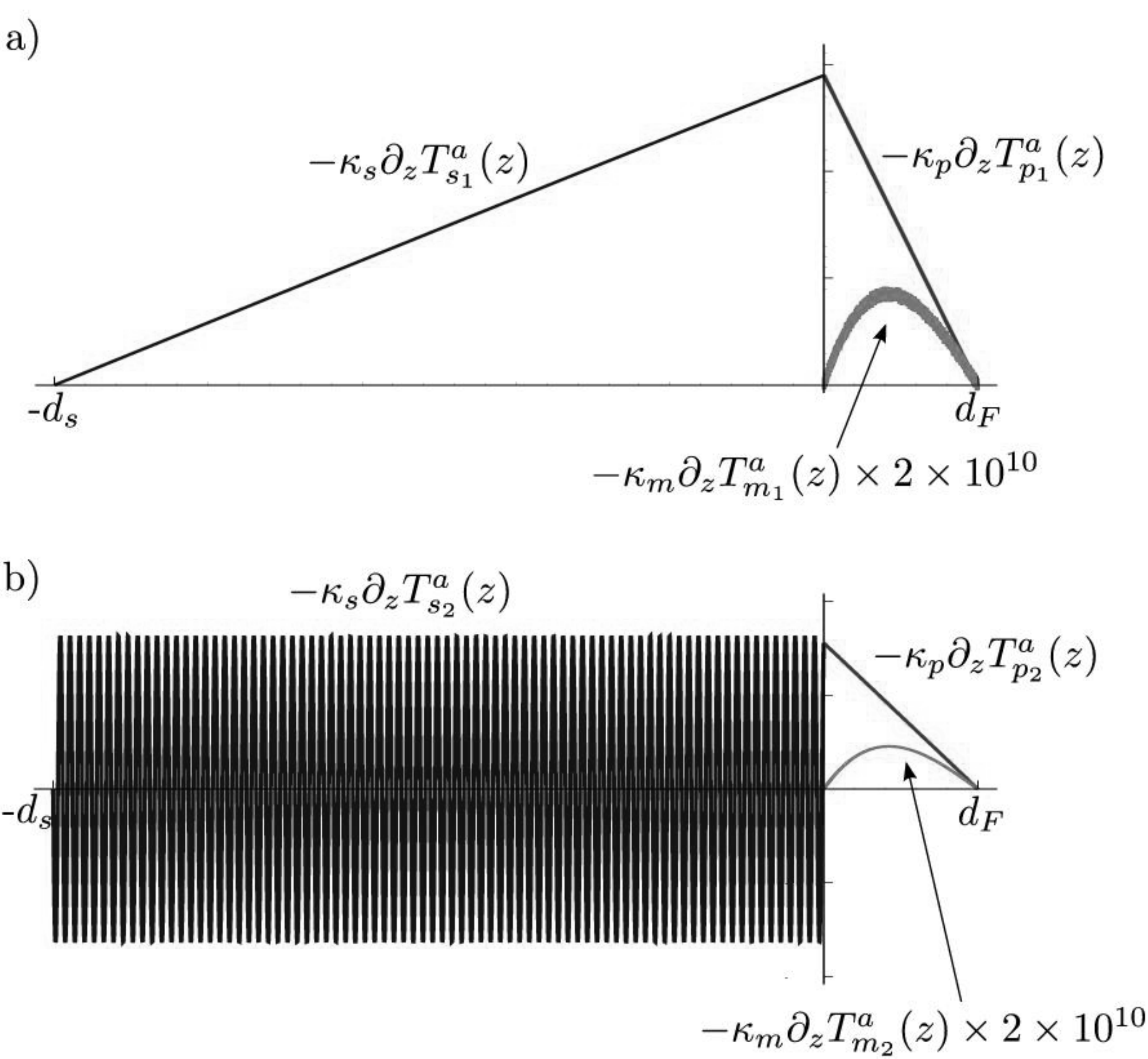}}
\caption
{The phonon and magnon heat fluxes (in arbitrary units) along $z$, for a given $x$, as a function of $z$, i.e., $-\kappa_{(s,p,m)} \partial_z T^a_{(s,p,m){_n}}(z)$, in the thermal equilibration modes with the two largest characteristic lengths.  The substrate occupies $z<0$ and the sample, with thickness magnified by 10$^{3}$, occupies $z>0$.  In the sample the magnon heat flux is nearly parabolic and the phonon heat flux is nearly linear.  In (a), where $n=1$, the heat flux in the substrate is nearly linear.  In (b), where $n=2$, the heat flux in the substrate has many oscillations because $\lambda_2^{\rm (2D)} \ll d_s$.  For both modes the sample is too thin for magnons to build up significant heat flux along $z$; in both (a) and (b) the magnon heat fluxes are magnified by $2 \times 10^{10}$.  }
\label{fig:jz2D}
\end{figure}

\begin{figure}[ht!]
\centerline{\includegraphics[width=0.46\textwidth]{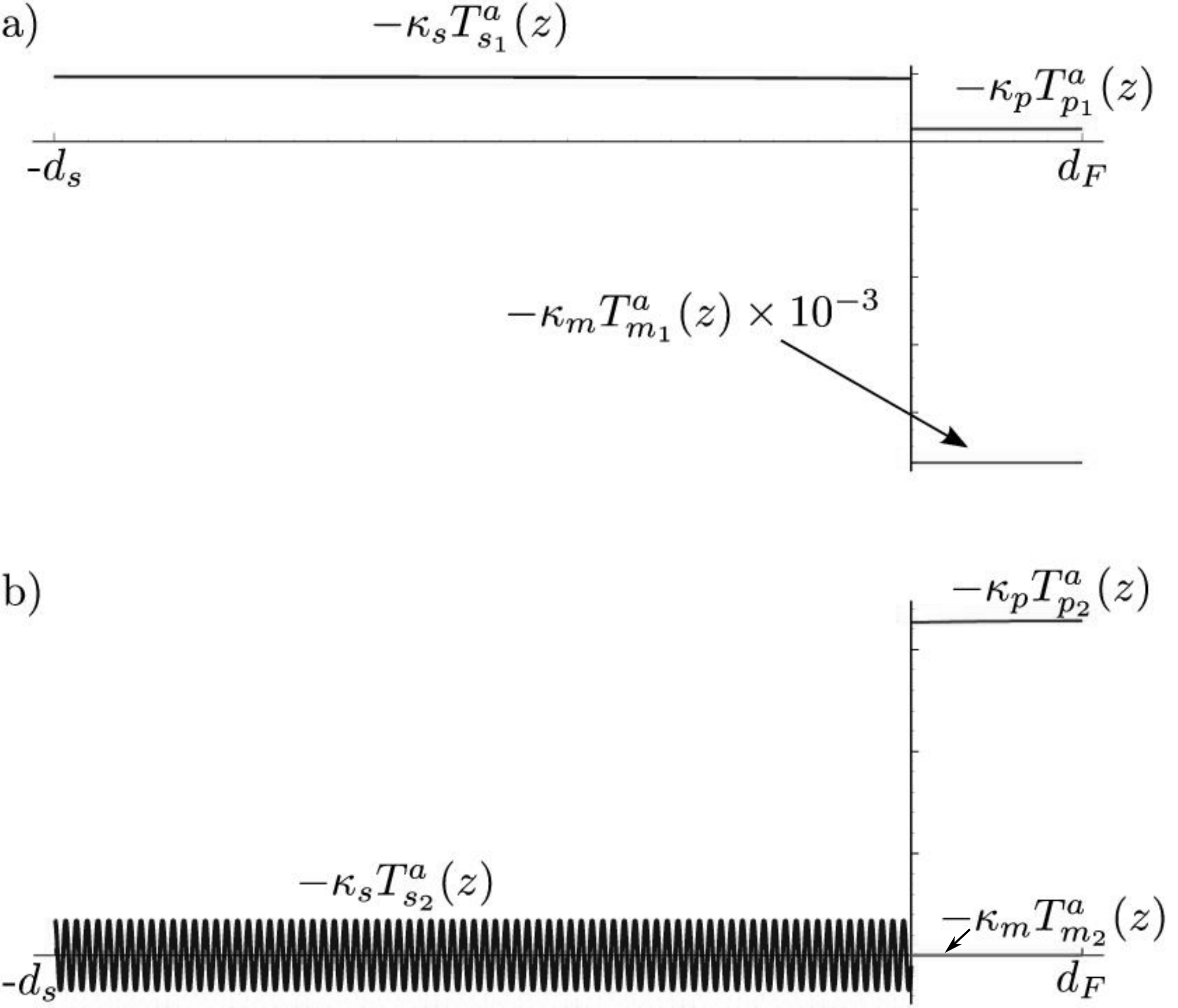}}
\caption
{The phonon and magnon heat fluxes (in arbitrary units) along $x$, for a given $x$, as a function of $z$, i.e., $-\kappa_{(s,p,m)} T^a_{(s,p,m){_n}}(z)$, in the thermal equilibration modes with the two largest characteristic lengths.  The substrate occupies $z<0$ and the sample, with thickness magnified by 10$^{3}$, occupies $z>0$.  In (a), where $n=1$, the magnon heat flux is multiplied by $10^{-3}$.  For the parameters of Table~\ref{tab:Jaworski}, (a) shows that along $x$ the heat flow for $n=1$ is carried by all three subsystems, with magnon heat flux opposing sample and substrate phonon heat flow, and (b) shows that along $x$ the heat flux for $n=2$ is carried mostly by the phonon subsystems, which oppose one another at the interface.  In (b), where $n=2$, the heat flux in the substrate has many oscillations because $\lambda_2^{\rm (2D)} \ll d_s$.  Although it is not obvious at this scale, each heat flux has some curvature.  
}
\label{fig:jx2D}
\vspace*{12pt}
\end{figure}

The subsystem contributions to heat flow along $z$ and along $x$ for the two modes associated with $q_{1}^{\rm (2D)}$ and $q_{2}^{\rm (2D)}$ are respectively shown in Figs.~\ref{fig:jz2D} and \ref{fig:jx2D}.  Fig.~\ref{fig:jx2D} explains the significant difference between $q_{2}^{\rm (1D)}$ and $q_{2}^{\rm (2D)}$; the 1D solutions $q_{1}^{\rm (1D)}$ and $q_{2}^{\rm (1D)}$ should apply for heat flux along $x$ uniform in $z$.  This holds for the $q_{1}^{\rm (2D)}$ mode in Fig.~\ref{fig:jx2D}a, whereas the $q_{2}^{\rm (2D)}$ displays significant curvature in Fig.~\ref{fig:jx2D}b.

\begin{figure*}[ht!]
\centerline{\includegraphics[width=0.90\textwidth]{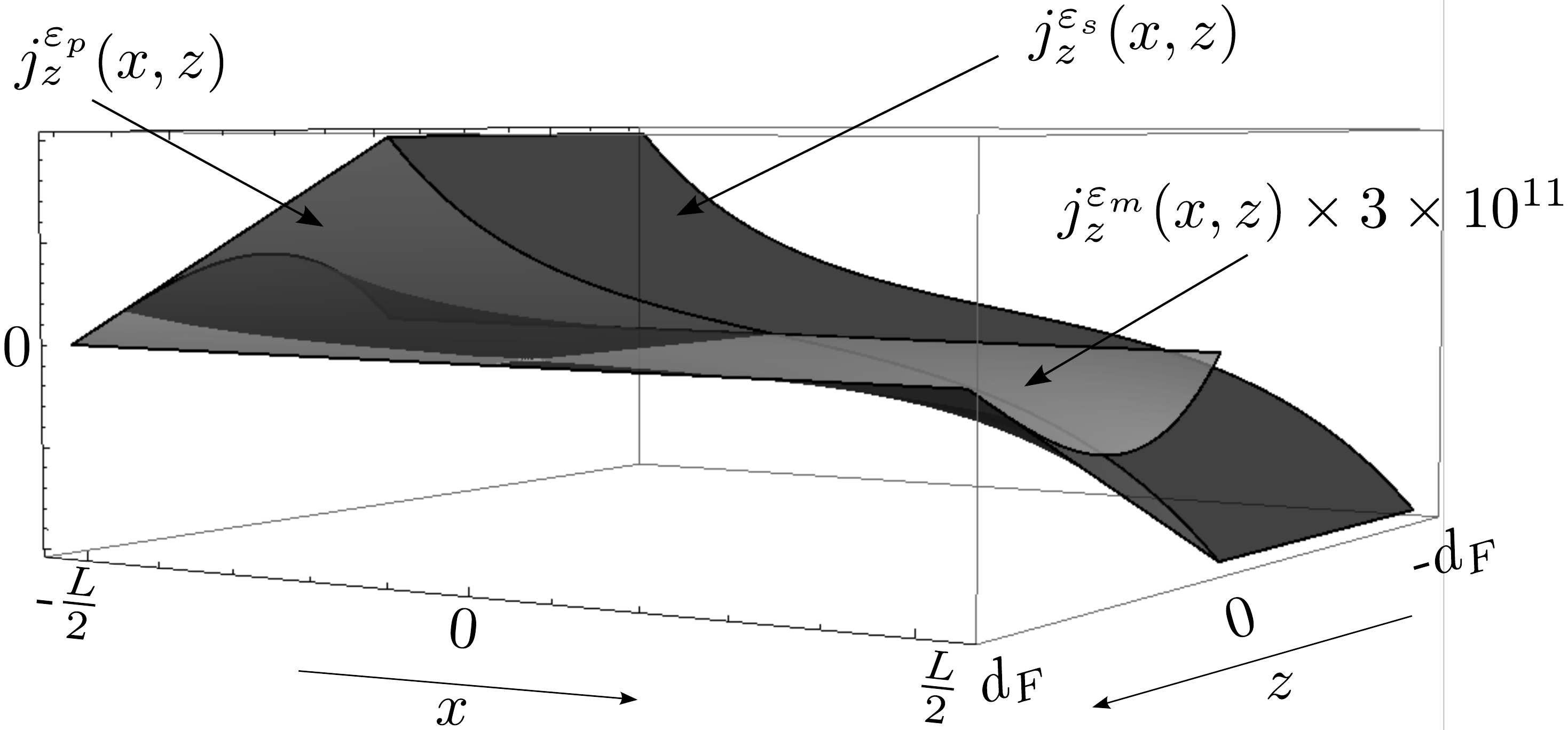}}
\caption
{The relative magnitudes of phonon and magnon heat flux along $z$ as a function of $x$ and $z$,
i.e., $j_{z}^{\varepsilon_{(s,p,m)}}$ 
in arbitrary units.  The substrate (only part of which is pictured) is at $z<0$ and the sample is at $z>0$. 
The sample magnon heat flux is magnified here by the factor $3 \times 10^{11}$; for the parameter values of Table~\ref{tab:Jaworski}, the sample is too thin for magnons to build up much heat flux along $z$.  
The profile of each subsystem's heat flux along $z$ varies as $\sin{(q_{n}x)}$. }
\label{fig:jz3D}
\end{figure*}

\subsection{Infinite Number of Inverse Lengths}
Other consistent solutions $q_{n\geq 3}^{\rm (2D)} > q_{2}^{\rm (2D)} > q_{1}^{\rm (2D)}$ can be found numerically.  
We are here searching for the normal modes associated with heat flow with the largest decay lengths, the larger $q$ (and therefore smaller $\lambda$) solutions are irrelevant to the current discussion.  We do, however, discuss the nature of these solutions.

\begin{figure}[ht!]
\centerline{\includegraphics[width=0.46\textwidth]{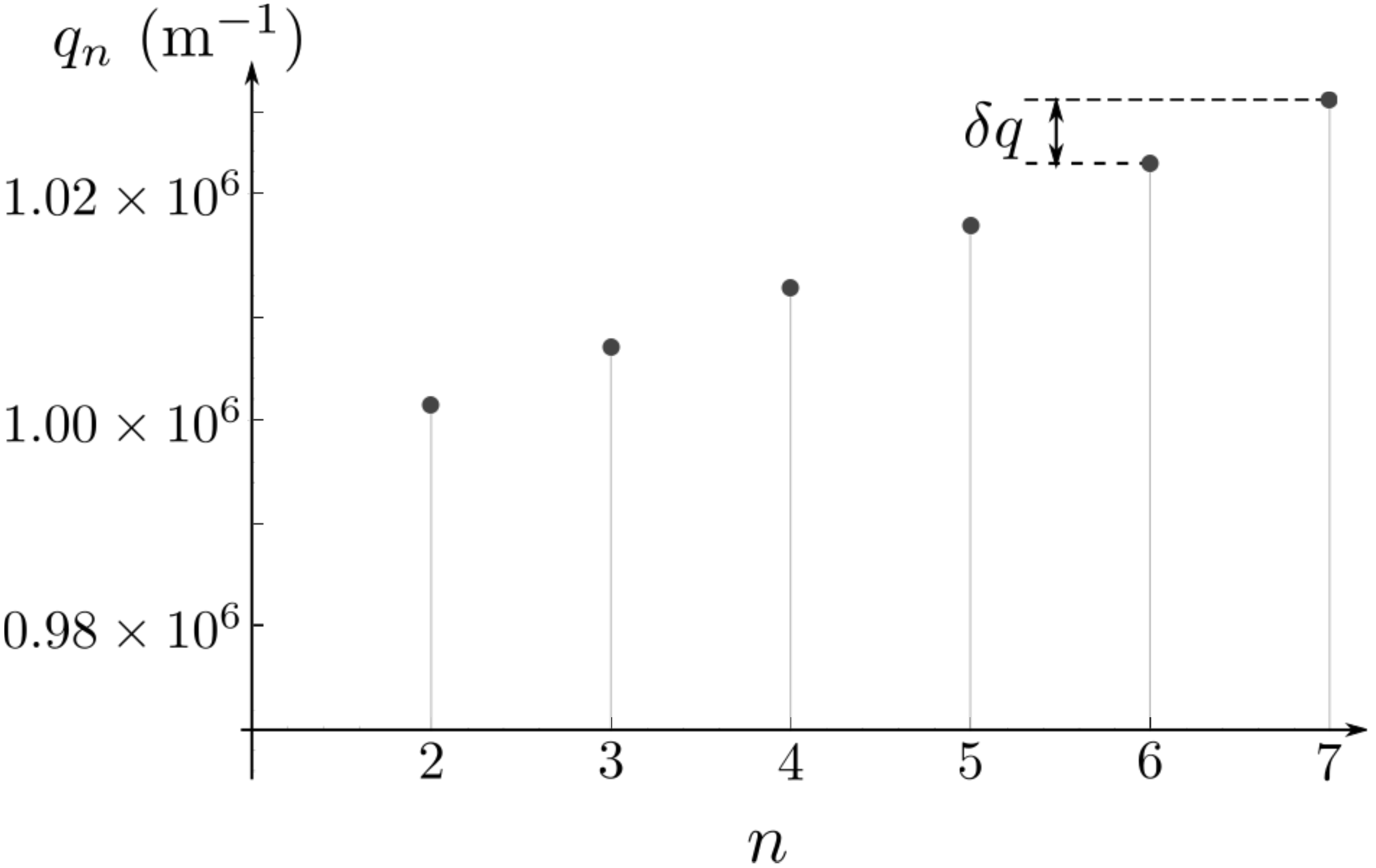}}
\caption{The inverse lengths $q_n$ for $n=2$ to $n=7$.  The inverse length $q_1$, which is not shown, is three orders of magnitude smaller than $q_2$.  The difference $\delta q$ between the inverse lengths of successive modes quickly approaches a value near $\pi/d_s \approx \pi/(d_s+ d_F)$, suggesting that the additional modes are associated with the physical geometry of the system.  }
\label{fig:InverseLengths}
\end{figure}

Figure~\ref{fig:InverseLengths} shows the magnitude of the seven smallest wavevectors (except $q_1$) versus the number of the solution $n$ (numbered by magnitude with $q_{n+1}^{\rm (2D)} > q_{n}^{\rm (2D)}$).  As $n$ grows, the difference $\delta q$ between the inverse lengths of successive modes approaches either $\pi/d_s$ or $\pi/(d_s+d_F)$; since $d_s \gg d_F$, it is difficult to distinguish which is the limiting quantity.  Thus, the higher solutions are associated with the geometry of the system.  We do not discuss them further.

Note that this numerical method, which searches for consistent values of $q$ by using trial values, might not obtain all solutions, no matter how exhaustive the list of trial values.  However, any missed modes are expected to have large $q$ and small $\lambda$, and thus are irrelevant to the current discussion.

\section{On the Measured Exponential Length}
\label{SSElength}

For the calculated maximum $\lambda_{mp}$ of Ref.~\onlinecite{Bauer1}, the present theory cannot account for the anomalously large length (on the order of $1$~mm) observed in the spin-Seebeck experiments.  On one hand, for the sample-substrate length $\lambda_{ps}$ to be on the order of 1~mm, with $\kappa_s \approx \kappa_p \sim 10^2$~W/m-K, $d_s \sim 100$~nm, and $d_F \sim 10$~nm, Eq.~\eqref{qmqsDef} gives an abnormally small thermal boundary conductance $h_K \sim 1$~W/m$^2$-K.  Although $h_K$ is not known for the particular combinations of materials used in Refs.~\onlinecite{UchidaPy,UchidaInsul,AwschMyers}, Fig.~34 of Ref.~\onlinecite{SwartzPohl} gives $h_K \approx10^{7}$~W/m$^2$-K (for Rh:Fe on Al$_2$O$_3$ at $T=50$~K). We do not expect that thermal matching between substrate and sample in the spin-Seebeck experiments to be considerably worse.  On the other hand, for the magnon-phonon length $\lambda_{mp}$ to be on the order of 1~mm, the mode coupling  term given by $R_{mp} R_{ps}$ in Eq.~\eqref{qSquare} would have to account for a large increase of $\lambda_{mp}$ (at least three-fold in the case of Permalloy.\citep{Bauer1})  Because spin-Seebeck experiments are carried out near room temperature\citep{UchidaPy,UchidaInsul} or at $T \geq 40$~K,\citep{AwschMyers} it is unlikely that the magnons carry a significant amount of the heat flux in the ferromagnet, i.e., it is likely that $\kappa_m \ll \kappa_p$.  Since the mode coupling term $R_{mp}$ is proportional to $\kappa_m/\kappa_p$, mode coupling is likely a weak effect. 

However, phonon-magnon drag, as proposed in Refs.~\onlinecite{AdachiDrag} and \onlinecite{JaworskiDrag}, or some other mechanism may explain a much longer $\lambda_{mp}$ than previously calculated.  
Hence, we have taken $\lambda_{mp}$ to be larger than estimated by Ref.~\onlinecite{Bauer1} (see Table~\ref{tab:Jaworski}). The results above show that, for such a large $\lambda_{mp}$, in the spin-Seebeck system we expect a thermal gradient along $z$ that varies as $\sinh{(x/\lambda)}$, resembling the $\Delta V_y$ measured by Ref.~\onlinecite{AwschMyers} (see its Fig.~2).   




\section{Relating Longitudinal Thermal Gradients to Transverse Voltage Differences}
\label{sec:SpinFlux}

The relation between the applied longitudinal temperature gradient and the transverse voltage difference is complicated, and worth  discussing.  So far we have shown that the applied longitudinal temperature gradient leads to a transverse (along $z$) temperature gradient in the sample -- the first of the three steps in Eq.~\eqref{LogicalFlow}, $\Delta T_x \xrightarrow{\rm Equil.} \partial_z T$.  In  Sec.~\ref{subsec:magnetochem} we show how to go from this tranverse temperature gradient to the accompanying transverse gradients of the magnetoelectrochemical potentials -- the second of the three steps in Eq.~\eqref{LogicalFlow}, $\partial_z T \xrightarrow{\rm SSE} \partial_z \bar{\mu}_{\uparrow,\downarrow}$ -- which are defined below.  Finally, in Sec.~\ref{subsec:spinhall} we show how to go from these transverse gradients (along $z$) of the magnetoelectrochemical potentials, via the up- and down- spin Hall conductivities, to the measured transverse (along $y$) voltage difference $\Delta V_y$ -- the third of the three steps in Eq.~\eqref{LogicalFlow}, $ \partial_z \bar{\mu}_{\uparrow,\downarrow} \xrightarrow{\rm ISHE} \Delta V_{y}$.  

We do not consider the use of platinum bars, which introduces a very complex geometry and is beyond the scope of the present work (and, as noted above, the effect has been observed with point contacts).  

\subsection{On Magnetoelectrochemical Potential, Temperature, and Spin Current}
\label{subsec:magnetochem}
By irreversible thermodynamics, the total spin flux (defined below as the difference of the number fluxes of up- and down-spin carriers), is driven both by gradients of temperature and of magnetoelectrochemical potentials.\cite{SearsSasHeating,JohnsonSilsbee}    
The magnetoelectrochemical potentials\footnote{Reference~\onlinecite{UchidaPy} appears to define the ``spin potential'' to be the difference of the up and down spin chemical potentials, and thus does not include the contribution from the magnetic disequilibrium associated with spin accumulation; this is different from the magnetoelectrochemical potential.} 
are defined by\cite{Saslow2007,SearsSasHeating} 
\begin{gather}
\bar{\mu}_{\uparrow,\downarrow}= \mu_{\uparrow,\downarrow} - e\phi \pm \frac{g \mu_B}{2} \vec{H}^* \cdot \hat{M}.
\label{mubar}
\end{gather}
Here, $\mu_{\uparrow}$ and $\mu_{\downarrow}$ are the chemical potentials of up- and down-spin electrons, $e$ is the electron charge, $\phi$ is electrical potential, $g$ is the electron $g$-factor, $\mu_B$ is the Bohr magneton, $\vec{H}^*$ is the effective magnetic field, and $\hat{M}$ is the direction of magnetization.  The field $\vec{H}^*$ is the difference between external magnetic fields and the internal fields, including the exchange and dipole contributions, and is defined so that $\vec{H}^*=0$ in equilibrium.  A more detailed discussion of  $\vec{H}^*$ is given in Ref.~\onlinecite{Saslow2007}.  


The up- and down- spin fluxes are primarily driven by the respective gradients $\bar{\mu}_{\uparrow}$
 and $\bar{\mu}_{\downarrow}$, but each has cross-terms\cite{JohnsonSilsbee,SearsSasHeating,Saslow2007} associated with the other potential, as well as with the temperature.  We thus write
 \begin{gather}
 j_i^{\uparrow} = - L_{\uparrow \varepsilon} \partial_i T_m - \frac{\sigma_{\uparrow}}{e^2} \partial_i \bar{\mu}_{\uparrow}  - L_{\uparrow \downarrow} \partial_i \bar{\mu}_{\downarrow} ,\label{jup}\\
 j_i^{\downarrow} = - L_{\downarrow \varepsilon} \partial_i T_m  - L_{\downarrow \uparrow} \partial_i \bar{\mu}_{\uparrow}  - \frac{\sigma_{\downarrow}}{e^2} \partial_i \bar{\mu}_{\downarrow} .\label{jdown}
 \end{gather}
Here, $\sigma_{\uparrow}$ and $\sigma_{\downarrow}$ are the respective bulk conductivities of up- and down-spins (generally not equal in a ferromagnet), and $L_{\uparrow \varepsilon}$ and $L_{\downarrow \varepsilon}$ (with units of m/K-s) are cross-term coefficients associating thermal gradients and individual spin-carrier currents (thus associated with both the electrical and spin currents).  By an Onsager relation\cite{Onsager31} $L_{\uparrow \downarrow}=L_{\downarrow \uparrow}$ (with units of m/J-s) are cross-term coefficients associating up and down spin currents with down and up magnetoelectrochemical gradients.  Typically $L_{\uparrow \downarrow}=L_{\downarrow \uparrow}$ are taken to be small, so that the terms $L_{\uparrow \downarrow} \partial_i \bar{\mu}_{\downarrow}$ and $L_{\downarrow \uparrow} \partial_i \bar{\mu}_{\uparrow}$ are negligible.

To calculate $j_i^{\uparrow}$ and $j_i^{\downarrow}$ everywhere, we employ their boundary conditions (that they have zero normal component at each sample boundary, which assumes no surface scattering) and their bulk equations, given for steady state by
\begin{align}
&\partial_i j_i^{\uparrow} = {\cal S}_{\uparrow \downarrow}, \quad
\partial_i j_i^{\downarrow} = {\cal S}_{\downarrow \uparrow}. \label{jupdownbulk}
\end{align}
For charge conservation, the up- and down-spin source terms ${\cal S}_{\uparrow \downarrow}$ and ${\cal S}_{\downarrow \uparrow}$ (which are proportional to $(\bar{\mu}_{\uparrow}-\bar{\mu}_{\downarrow})/\tau_{sf}$, where $\tau_{sf}$ is a characteristic spin-flip time\cite{Saslow2007,SearsSasHeating}) are equal and opposite.  
%
Substitution from Eqs.~\eqref{jup} and \eqref{jdown} into Eq.~\eqref{jupdownbulk} gives two equations for two unknowns, $\bar{\mu}_{\uparrow}$ and $\bar{\mu}_{\downarrow}$.  Because the temperatures are shown above to vary as $\sinh{(x/\lambda)}$, then $\bar{\mu}_{\uparrow}$ and $\bar{\mu}_{\downarrow}$ also vary as $\sinh{(x/\lambda)}$.  



\subsection{On the Spin Hall Effect}
\label{subsec:spinhall}
We now discuss how to go from $\partial_z \bar{\mu}_{\uparrow}$ and $\partial_z \bar{\mu}_{\downarrow}$ to the measured voltage difference along $y$, i.e., $\Delta V_y$. 
We work by analogy to the Hall effect, which occurs when an electric flux $\vec{J}$ is driven through a conductor in the presence of a magnetic field $\vec{B}'$ that is perpendicular to the current. 

Consider a conductor of width $w$ along $y$. 
Let the electric current be driven along $z$ by an applied electric field $E_{z}$, so that charge carriers have a velocity $v_{z}$.  With an applied magnetic field $(B'_x,0,0)$, a Lorentz force then drives the charge carriers along $y$, so that charges of opposite signs accumulate at the edges.  
The Lorentz-force-induced current is given by $J'_{y} =  \sigma v_z B'_x$.  In the steady state, there is no flow along $y$, so an electric field $E_{y}$ develops to oppose the Lorentz-induced current  along $y$.   The total charge flux along ${y}$ is given by
\begin{gather}
J_y = 0 =\sigma \left(E_y + v_z B'_x \right) .  
\end{gather}
The so-called Hall field $E_{y}$ thus is given by
\begin{gather}
E_{y} = -  v_z B'_x =   \frac{J_z B'_x}{n e},
\end{gather}
where we have used $\vec{J} = - n e \vec{v}$, and $n$ and $-e$ are the respective concentration and the charge of the charge carriers.  The Hall voltage is $\Delta V_y = E_y w$. 

Thus, the Hall effect relates an applied electric current to a measured transverse electrical potential difference. 
In contrast, the Spin Hall effect (SHE) relates an applied electric current to transverse differences in the magnetoelectrochemical potentials, and the inverse Spin Hall effect (ISHE) relates an applied spin current to a transverse difference in electrical potential (see, for example, Refs.~\onlinecite{DyakonovPerel,HirschSHE,ZhangSHE,SinovaISHE,SaitohISHE,ChudnovskySHE}). 
For the SHE and ISHE 
there are fluxes of charge carriers with both up- and down- spin.  Instead of the action of Lorentz force in the Hall effect, for the SHE there are forces due to the spin-orbit interaction, whose effect enters via non-zero up- and down- spin Hall conductivities $\sigma_{sH{_\uparrow}}$ and $\sigma_{sH{_\uparrow}}$.  (Thus the effect of the spin-orbit interaction is taken to be a perturbation.)  Instead of the electric field $E_{y} = -\partial_y \phi$, the spin-orbit force is associated with $ -\partial_y \bar{\mu}_{\uparrow}$ and $ -\partial_y \bar{\mu}_{\downarrow}$. 
We take the contributions to the number fluxes  along $y$ of the up- and down- spin carriers by this spin-orbit force\footnote{To eliminate the applied magnetic field $B_x$, rather than the spin-orbit interaction, as the source of the deflection of the spin carriers leading to $\Delta V_y$, Ref.~\onlinecite{UchidaInsul} shows that $\Delta V_y=0$ when Cu bars (with weak spin-orbit interaction) are used for detection, whereas $\Delta V_y \neq 0$ for Pt bars (with much stronger spin-orbit interaction).}  to be given by 
\begin{gather}
 j_y^{sH{\uparrow}} = \frac{\sigma_{sH{_\uparrow}}}{e} \partial_z \bar{\mu}_{\uparrow}   ,\quad
j_y^{sH{\downarrow}} = \frac{\sigma_{sH{_\downarrow}}}{e} \partial_z \bar{\mu}_{\downarrow},
\label{SHcont}
\end{gather}
The number fluxes  along $y$ of the up- and down- spin carriers can be written as
\begin{align}
&j_y^{\uparrow} = -\frac{\sigma_{\uparrow}}{e} \partial_y \bar{\mu}_{\uparrow} +\frac{\sigma_{sH{_\uparrow}}}{e} \partial_z \bar{\mu}_{\uparrow} ,\label{jupSH}\\
&j_y^{\downarrow} = -\frac{\sigma_{\downarrow}}{e} \partial_y \bar{\mu}_{\downarrow} +\frac{\sigma_{sH{_\downarrow}}}{e} \partial_z \bar{\mu}_{\downarrow} .  
\label{jdownSH}
\end{align}

For no charge current along $y$, the sum $j_y^{\uparrow} + j_y^{\downarrow}=0$.  We also assume no bulk spin current along $y$, so $j_y^{\uparrow} - j_y^{\downarrow}=0$.  
Thus, we take $j_y^{\uparrow}  =0$ and $ j_y^{\downarrow}=0$, so that Eqs.~\eqref{jupSH} and \eqref{jdownSH} give
\begin{gather}
\partial_y \bar{\mu}_{\uparrow} = \frac{ \sigma_{sH{_\uparrow}} }{\sigma_{\uparrow}} \partial_z \bar{\mu}_{\uparrow}, \quad 
\partial_y \bar{\mu}_{\downarrow} = \frac{ \sigma_{sH{_\downarrow}} }{\sigma_{\downarrow}} \partial_z \bar{\mu}_{\downarrow}.
\label{SHE}
\end{gather}
The known sources $\partial_z \bar{\mu}_{\uparrow}$ and $\partial_z \bar{\mu}_{\downarrow}$ on the right-hand-sides (RHS) of Eq.~\eqref{SHE} are uniform in $y$.

To write the magnetoelectrochemical potential in terms of the concentrations of up- and down-spins and the electric potential, we linearize the chemical potentials and the effective magnetic field term as
\begin{gather}
\delta \mu_{\uparrow,\downarrow} =  \frac{\partial \mu_{\uparrow,\downarrow}}{\partial n_{\uparrow,\downarrow}} \delta n_{\uparrow,\downarrow}, \quad \delta \vec{H}^* \cdot \hat{M} = \frac{\mu_0 \mu_B}{\chi} \left( \delta n_{\uparrow} - \delta n_{\downarrow} \right),
\label{muHlin}
\end{gather}
where $\delta$ denotes deviations from equilibrium, $\mu_0$ is the permeability of free space, and $\chi$ is the magnetic susceptibility.
Then  Eq.~\eqref{mubar} gives
\begin{gather}
\delta \bar{\mu}_{\uparrow,\downarrow}= \frac{\partial \mu_{\uparrow,\downarrow}}{\partial n_{\uparrow,\downarrow}} \delta n_{\uparrow,\downarrow} - e \delta \phi \pm \frac{g \mu_0 \mu_B^2}{2 \chi} \left( \delta n_{\uparrow} - \delta n_{\downarrow} \right).
\label{mubarlin}
\end{gather}

With ${\partial \mu_{\uparrow,\downarrow}}/{\partial n_{\uparrow,\downarrow}}$ uniform in $y$, substitution of Eq.~\eqref{mubarlin} into the left-hand-sides (LHS) of Eq.~\eqref{SHE} gives
\begin{align}
\frac{\partial \mu_{\uparrow}}{\partial n_{\uparrow}} \partial_y \delta n_{\uparrow} - e \partial_y \delta \phi + \frac{g \mu_0 \mu_B^2}{2 \chi} \left( \partial_y \delta n_{\uparrow} - \partial_y \delta n_{\downarrow} \right) = \frac{ \sigma_{sH{_\uparrow}} }{\sigma_{\uparrow}} \partial_z \bar{\mu}_{\uparrow},\label{3up}\\
\frac{\partial \mu_{\downarrow}}{\partial n_{\downarrow}} \partial_y \delta n_{\downarrow} - e \partial_y \delta \phi - \frac{g \mu_0 \mu_B^2}{2 \chi} \left( \partial_y \delta n_{\uparrow} - \partial_y \delta n_{\downarrow} \right) = \frac{ \sigma_{sH{_\downarrow}} }{\sigma_{\downarrow}} \partial_z \bar{\mu}_{\downarrow}.\label{3down}
\end{align}
With the RHS of Eqs.~\eqref{3up} and \eqref{3down} known, they are two equations for the three unknowns $\delta n_{\uparrow}$, $\delta n_{\downarrow}$, and $\delta \phi$.  A third relation is provided by Gauss's Law:
\begin{gather}
\partial_y^2 \delta \phi = -\frac{e}{\varepsilon_0 \varepsilon} \left(\delta n_{\uparrow} + \delta n_{\downarrow} \right),
\label{Gauss}
\end{gather}
where $\varepsilon_0$ and $\varepsilon$ are the permittivity of free space and the relative permittivity.  Solving Eqs.~\eqref{3up}-\eqref{Gauss} gives $\delta n_{\uparrow}$, $\delta n_{\downarrow}$, and $\delta \phi$, the last of which is related to the measured voltage by $\Delta V_y = \int_{-w/2}^{w/2} dy \delta \phi$.  We now discuss the solution.

It is consistent to take $\delta n_{\uparrow} = - \delta n_{\downarrow}$, i.e., local electroneutrality.\footnote{Because Ref.~\onlinecite{UchidaPy} takes the ``spin potential'' to be the difference of up- and down-spin chemical potentials, it neglects the effective field $H^*$ and therefore by Eq.~\eqref{muHlin} assumes no spin accumulation.  The present work, however, permits spin accumulation. }   
(Equations~\eqref{3up}-\eqref{Gauss} then give that  $\partial_y \delta \phi$ and $\partial_y \delta n_{\uparrow}$ are uniform in $y$.)  Equations~\eqref{3up} and \eqref{3down} can then be 
solved for $\partial_y \delta n_{\uparrow}$ and $\partial_y \delta \phi$. 
Defining the dimensionless ratio
\begin{gather}
R_{\mu} \equiv  \frac{\displaystyle \frac{\partial \mu_{\uparrow}}{\partial n_{\uparrow}} -\frac{\partial \mu_{\downarrow}}{\partial n_{\downarrow}}  }
{\displaystyle  \frac{\partial \mu_{\uparrow}}{\partial n_{\uparrow}} +\frac{\partial \mu_{\downarrow}}{\partial n_{\downarrow}} + \frac{2 g \mu_0 \mu_B^2}{\chi} },
\end{gather}
we have\begin{gather}
\partial_y \delta n_{\uparrow}  =  R_{\mu} \left( \frac{\partial \mu_{\uparrow}}{\partial n_{\uparrow}} -\frac{\partial \mu_{\downarrow}}{\partial n_{\downarrow}}  \right)^{-1} \left( \frac{ \sigma_{sH{_\uparrow}} }{\sigma_{\uparrow}} \partial_z \bar{\mu}_{\uparrow} - \frac{ \sigma_{sH{_\downarrow}} }{\sigma_{\downarrow}} \partial_z \bar{\mu}_{\downarrow}\right).
\label{deltanup}\\
\partial_y \delta \phi = -\left(\frac{1-R_{\mu}}{2e} \right) \frac{ \sigma_{sH{_\uparrow}} }{\sigma_{\uparrow}} \partial_z \bar{\mu}_{\uparrow} - \left(\frac{1+R_{\mu}}{2e} \right)\frac{ \sigma_{sH{_\downarrow}} }{\sigma_{\downarrow}} \partial_z \bar{\mu}_{\downarrow}.
 \label{deltaphi}
\end{gather}
With $\Delta V_y = \int_{-w/2}^{w/2} dy \delta \phi$, integration of Eq.~\eqref{deltaphi} over $y$ across the width of the sample then gives
\begin{gather}
\Delta V_y = \frac{w}{2e} \left[\left(R_{\mu} -1 \right) \frac{ \sigma_{sH{_\uparrow}} }{\sigma_{\uparrow}} \partial_z \bar{\mu}_{\uparrow} - \left(R_{\mu} +1 \right)\frac{ \sigma_{sH{_\downarrow}} }{\sigma_{\downarrow}} \partial_z \bar{\mu}_{\downarrow} \right],
\label{Vy}
\end{gather}
where we have employed the uniformity of $\partial_z \bar{\mu}_{\uparrow}$ and $\partial_z \bar{\mu}_{\downarrow}$ along $y$.  As discussed above, $\bar{\mu}_{\uparrow}$ and $\bar{\mu}_{\downarrow}$ vary as $\sinh{(x/\lambda)}$, thus Eq.~\eqref{Vy} predicts $\Delta V_y \sim \sinh{(x/\lambda)}$.


The present work shows that the relation between $\Delta V_y$ and $\Delta T_x$ is very complicated, and suggests that a direct relation $\Delta V_y \sim S_S \Delta T_x$ (see, e.g., Ref.~\onlinecite{UchidaPy}) is correct, but may not be quantitatively useful.  However, the present work does support such a qualitative analysis, where the applied thermal gradient along $x$ leads, via the spin-Seebeck effect, to spin carrier fluxes along $z$, which in turn produce the measured voltage difference $\Delta V_{y}$ along $y$.

When surface scattering occurs, the present analysis would become much more complex; see Landauer and Swanson\cite{Landauer53} for the ordinary Hall effect. 


\section{Summary and Conclusion}
\label{sec:Conclusion}

The present work finds the detailed temperature profile for the spin-Seebeck system, including both sample and substrate, when a temperature difference $\Delta T_x$ is applied along $x$.   For 1D heat flow (only along $x$) we find that the temperature contains a part varying as $\sinh{(x/\lambda)}$, for each of two characteristic lengths ($\lambda_{ps}$ and $\lambda_{mp}$), one of which may correspond to the observed decay length of $\Delta V_y$.  Equations~\eqref{qSquare} and \eqref{qmqsDef} show that quadrupling the thickness of both the sample and substrate should approximately double these lengths.  
Polishing (roughening) the substrate before depositing the sample should increase (decrease) $h_K$, and thus decrease (increase) $\lambda_{ps}$.   If $\lambda_{ps}$ corresponds to the observed exponential decay length, measurements on a series of samples with increasingly rough sample/substrate interfaces should reveal this dependence.  
Further, changing the coupling factor between the modes (by changing $\kappa_m/\kappa_p$ or $d_s \kappa_s/d_F \kappa_p$) modifies both lengths -- increasing either increases the larger length, which likely corresponds to the measured decay length of $\Delta V_y$.  

For 2D heat flow (along both $x$ and $z$), we also find that the temperature and thermal gradients along $z$ in the spin-Seebeck system vary as $\sinh{(x/\lambda)}$, and find a complicated sinusoidal and exponential profile along $z$ for the thermal gradients, with an infinite number of characteristic lengths, which we study numerically.  The longest of these corresponds to the longest 1D length.  The second longest length is a geometry-modified version of the other 1D length.  Further lengths are largely due to the geometry.   

We show how the thermal gradient along $x$ leads to the measured $\Delta V_{y}$.  The thermal gradient along $x$ leads to a thermal gradient along $z$, which then drives up- and down- spin currents along $z$ (the spin-Seebeck effect), and is accompanied by gradients along $z$ of the magnetoelectrochemical potentials.  These magnetoelectrochemical potential gradients along $z$ then produce the measured $\Delta V_{y}$, via the inverse Spin Hall effect (due to a nonzero spin-orbit interaction that leads to spin-Hall conductivities).  


\section{Acknowledgements}
We would like to acknowledge V. Pokrovsky for valuable conversations, and the support of the Department of Energy through grant DE-FG02-06ER46278.

\bibliographystyle{apsrev4-1}
\bibliography{IrrThBib}{}

\appendix

\section{Bulk and Boundary Conditions for Heat Flow along $x$ and $z$}
\label{App:Details}
With Eqs.~\eqref{T0alpha} and \eqref{T0alphas} relating the linear terms in temperature, there are $2+10N$ unknowns in Eqs.~\eqref{jxGen}, \eqref{jzGen}, \eqref{Gs} and \eqref{Ggen}: one $T_0$, one $\alpha$, and $10N$ amplitudes given by $A_{s{_n}}^{(1,2)}$, $A_{p{_n}}^{(1,2,3,4)}$, and $A_{m{_n}}^{(1,2,3,4)}$.  This section details the bulk and boundary conditions on heat flux that give these unknowns.

\subsection{Bulk Conditions}

By matching coefficients of like terms, substitution of Eqs.~\eqref{Tgen} and \eqref{Ggen} into Eq.~\eqref{d2TpTm2} gives
\begin{gather}
A_{m{_n}}^{(1)} = - \frac{\kappa_p}{\kappa_m} A_{p{_n}}^{(1)},\quad A_{m{_n}}^{(2)} = - \frac{\kappa_p}{\kappa_m} A_{p{_n}}^{(2)},\notag\\
A_{m{_n}}^{(3)} = A_{p{_n}}^{(3)} , \quad A_{m{_n}}^{(4)} = A_{p{_n}}^{(4)} .
\label{AmAp}
\end{gather}
Since each of the above relations is a single condition for each mode $n=1,2,...N$, then Eq.~\eqref{AmAp} gives $4N$ conditions.

\subsection{Boundary Conditions}

\subsubsection{Boundary Conditions on Heat Flux along $z$}

There are a further $5N$ conditions given by the boundary conditions on the heat flux along $z$ for the various subsystems at $z=-d_s$, $z=0$, and $z=d_F$.  They are given above as Eqs.~\eqref{jzmATdf}-\eqref{jzsATds}, and any two of Eqs.~\eqref{jzsAT0}-\eqref{jzmAT0} with the third implicitly guaranteed by energy conservation.  

\subsubsection{Boundary Conditions on Heat Flux along $x$}
Two further conditions that constrain the homogeneous temperature coefficients, $T_0$ and $\alpha$, come from the temperatures of the heater and the heat sink.  
%
The remaining conditions on heat flux along $x$ are not obvious.  

With the heater and heat sink each in contact only with the substrate, we take the boundary conditions in the $x$-direction on each energy flux $j_x^{\varepsilon}$ are symmetric (we employ this above in taking $T^{b}_{(s,p,m)}(z)=0$).  This precludes permitting the heat flux input by the heater to have a different profile in $z$ than the heat flux output to the heat sink.  However, as stated above, we are only treating the region far enough away from the heaters that the details of heat flux entering and leaving at $x=\pm L/2$ are irrelevant.  Only a full solution with an infinite sum over inverse lengths $q_n$ can treat the specifics of the interfacial input, and it is beyond the scope of this work to solve for infinite inverse lengths.  Thus, we can not apply boundary conditions precisely at $x=\pm L/2$.  

We make the following approximation: at $x=\pm L/2 \mp \ell_S$, where $\ell_S$ is just far enough away from the heater/heat sink that the details of the input heat flux are irrelevant, we take $\partial_x T_p=0$ and $\partial_x T_m=0$.  
We take the heaters to be in contact only with the substrate, and assume that a  significant amount of heat does not seep into the sample over the distance $\ell_S$.  Explicitly,
\begin{gather}
\partial_x T_m(x=-L/2 + \ell_S)=0,\label{xCond1}\\
\partial_x T_p(x=-L/2 + \ell_S)=0.\label{xCond2}
\end{gather}
Recall that we take heat flux (and therefore $\partial_x T$) to be symmetric about $x=0$, so that the conditions at $x=+L/2-\ell_S$ are not independent.  
Although it is not obvious, Eqs.~\eqref{xCond1} and \eqref{xCond2} give $N$ conditions, which relate the amplitudes of each of the $N$ surface modes to the others.  

Thus, for the 2 + 10$N$ unknowns in the substrate phonon, sample phonon, and sample magnon temperatures associated with heat flow along both $x$ and $z$, Eqs.~\eqref{jzmATdf}-\eqref{jzmAT0} and \eqref{AmAp}-\eqref{xCond2} give 2 + 10$N$ conditions, and there are no free parameters.

\end{document}